\pdfoutput=1
\documentclass[12pt,a4paper]{article}

\usepackage{ifthen} 
\newboolean{pdflatex}
\setboolean{pdflatex}{true} 

\newboolean{articletitles}
\setboolean{articletitles}{true} 

\newboolean{uprightparticles}
\setboolean{uprightparticles}{false} 

\def\paperauthors{LHCb collaboration} 
\def\paperasciititle{First observation of the decay B0->D0D0barKpi} 
\def\papertitle{First observation of the decay \sigdecay} 
\def\paperkeywords{{High Energy Physics}, {LHCb}} 
\def\papercopyright{\the\year\ CERN for the benefit of the LHCb collaboration} 
\def\paperlicence{CC BY 4.0 licence}
\def\paperlicenceurl{https://creativecommons.org/licenses/by/4.0/}

\usepackage[top=1in, bottom=1.25in, left=1in, right=1in]{geometry}

\usepackage{microtype}
\usepackage{lineno}  
\usepackage{xspace} 
\usepackage{caption} 

\usepackage{graphicx}  
\usepackage{color}
\usepackage{colortbl}
\graphicspath{{./figs/}} 

\usepackage{amsmath} 
\usepackage{amssymb}
\usepackage{amsfonts}
\usepackage{upgreek} 

\newcommand*\patchAmsMathEnvironmentForLineno[1]{%
\expandafter\let\csname old#1\expandafter\endcsname\csname #1\endcsname
\expandafter\let\csname oldend#1\expandafter\endcsname\csname
end#1\endcsname
 \renewenvironment{#1}%
   {\linenomath\csname old#1\endcsname}%
   {\csname oldend#1\endcsname\endlinenomath}%
}
\newcommand*\patchBothAmsMathEnvironmentsForLineno[1]{%
  \patchAmsMathEnvironmentForLineno{#1}%
  \patchAmsMathEnvironmentForLineno{#1*}%
}
\AtBeginDocument{%
\patchBothAmsMathEnvironmentsForLineno{equation}%
\patchBothAmsMathEnvironmentsForLineno{align}%
\patchBothAmsMathEnvironmentsForLineno{flalign}%
\patchBothAmsMathEnvironmentsForLineno{alignat}%
\patchBothAmsMathEnvironmentsForLineno{gather}%
\patchBothAmsMathEnvironmentsForLineno{multline}%
\patchBothAmsMathEnvironmentsForLineno{eqnarray}%
}

\usepackage{hyperxmp}

\usepackage[pdftex,
            pdfauthor={\paperauthors},
            pdftitle={\paperasciititle},
            pdfkeywords={\paperkeywords},
            pdfcopyright={Copyright (C) \papercopyright},
            pdflicenseurl={\paperlicenceurl}]{hyperref}

\usepackage[colorinlistoftodos,textsize=scriptsize]{todonotes}

\usepackage[bottom,flushmargin,hang,multiple]{footmisc}

\usepackage[all]{hypcap} 

\usepackage{xspace} 
\usepackage{upgreek}

\def\lhcb   {\mbox{LHCb}\xspace}

\def\MagUp {\mbox{\em Mag\kern -0.05em Up}\xspace}

\ifthenelse{\boolean{uprightparticles}}%
{

 \def\Ppi         {\ensuremath{\uppi}\xspace}

 \def\PDelta      {\ensuremath{\Delta}\xspace}                 
 \def\PXi         {\ensuremath{\Xi}\xspace}                 
 \def\PLambda     {\ensuremath{\Lambda}\xspace}                 
 \def\PSigma      {\ensuremath{\Sigma}\xspace}                 
 \def\POmega      {\ensuremath{\Omega}\xspace}                 
 \def\PUpsilon    {\ensuremath{\Upsilon}\xspace}

 \def\PB      {\ensuremath{\mathrm{B}}\xspace}                 
                  
 \def\PD      {\ensuremath{\mathrm{D}}\xspace}

 \def\PK      {\ensuremath{\mathrm{K}}\xspace}

 \def\Pb      {\ensuremath{\mathrm{b}}\xspace}                 
 \def\Pc      {\ensuremath{\mathrm{c}}\xspace}

 \def\Pi      {\ensuremath{\mathrm{i}}\xspace}

 \def\Pp      {\ensuremath{\mathrm{p}}\xspace}

 \def\Ps      {\ensuremath{\mathrm{s}}\xspace}

 \def\thebaroffset{0.0em}
}
{

 \def\Ppi         {\ensuremath{\pi}\xspace}

 \mathchardef\PDelta="7101
 \mathchardef\PXi="7104
 \mathchardef\PLambda="7103
 \mathchardef\PSigma="7106
 \mathchardef\POmega="710A
 \mathchardef\PUpsilon="7107
                  
 \def\PB      {\ensuremath{B}\xspace}                 
                  
 \def\PD      {\ensuremath{D}\xspace}

 \def\PK      {\ensuremath{K}\xspace}

 \def\Pb      {\ensuremath{b}\xspace}                 
 \def\Pc      {\ensuremath{c}\xspace}

 \def\Pi      {\ensuremath{i}\xspace}

 \def\Pp      {\ensuremath{p}\xspace}

 \def\Ps      {\ensuremath{s}\xspace}

 \def\thebaroffset{0.18em}
}
\newcommand{\offsetoverline}[2][\thebaroffset]{\kern #1\overline{\kern -#1 #2}}%

\makeatletter
\ifcase \@ptsize \relax
  \newcommand{\miniscule}{\@setfontsize\miniscule{4}{5}}
\or
  \newcommand{\miniscule}{\@setfontsize\miniscule{5}{6}}
\or
  \newcommand{\miniscule}{\@setfontsize\miniscule{5}{6}}
\fi
\makeatother

\DeclareRobustCommand{\optbar}[1]{\shortstack{{\miniscule (\rule[.5ex]{1.25em}{.18mm})}
  \\ [-.7ex] $#1$}}

\def\squark    {{\ensuremath{\Ps}}\xspace}

\def\cquark    {{\ensuremath{\Pc}}\xspace}

\def\bquark    {{\ensuremath{\Pb}}\xspace}

\def\pion   {{\ensuremath{\Ppi}}\xspace}

\def\pip    {{\ensuremath{\pion^+}}\xspace}
\def\pim    {{\ensuremath{\pion^-}}\xspace}

\def\kaon    {{\ensuremath{\PK}}\xspace}

\def\KorKbar {\kern \thebaroffset\optbar{\kern -\thebaroffset \PK}{}\xspace}

\def\Kp      {{\ensuremath{\kaon^+}}\xspace}
\def\Km      {{\ensuremath{\kaon^-}}\xspace}

\def\Kstarz  {{\ensuremath{\kaon^{*0}}}\xspace}

\def\Kstar   {{\ensuremath{\kaon^*}}\xspace}

\def\Dbar    {{\ensuremath{\offsetoverline{\PD}}}\xspace}
\def\D       {{\ensuremath{\PD}}\xspace}
\def\Db      {{\ensuremath{\Dbar}}\xspace}
\def\DorDbar {\kern \thebaroffset\optbar{\kern -\thebaroffset \PD}\xspace}
\def\Dz      {{\ensuremath{\D^0}}\xspace}
\def\Dzb     {{\ensuremath{\Dbar{}^0}}\xspace}
\def\Dp      {{\ensuremath{\D^+}}\xspace}
\def\Dm      {{\ensuremath{\D^-}}\xspace}

\def\DpDm    {\ensuremath{\Dp {\kern -0.16em \Dm}}\xspace}

\def\Dstarz  {{\ensuremath{\D^{*0}}}\xspace}
\def\Dstarzb {{\ensuremath{\Dbar{}^{*0}}}\xspace}

\def\Dstarm  {{\ensuremath{\D^{*-}}}\xspace}

\def\B       {{\ensuremath{\PB}}\xspace}

\def\BorBbar {\kern \thebaroffset\optbar{\kern -\thebaroffset \PB}\xspace}
\def\Bz      {{\ensuremath{\B^0}}\xspace}

\def\Bd      {{\ensuremath{\B^0}}\xspace}

\def\BdorBdbar {\kern \thebaroffset\optbar{\kern -\thebaroffset \Bd}\xspace}
\def\Bu      {{\ensuremath{\B^+}}\xspace}

\def\Bp      {{\ensuremath{\Bu}}\xspace}

\def\Bs      {{\ensuremath{\B^0_\squark}}\xspace}

\def\BsorBsbar {\kern \thebaroffset\optbar{\kern -\thebaroffset \Bs}\xspace}

\def\Y#1S{\ensuremath{\PUpsilon{(#1S)}}\xspace}

\def\proton      {{\ensuremath{\Pp}}\xspace}
\def\antiproton  {{\ensuremath{\overline \proton}}\xspace}

\def\Lbar        {{\ensuremath{\offsetoverline{\PLambda}}}\xspace}
\def\LorLbar     {\kern \thebaroffset\optbar{\kern -\thebaroffset \PLambda}\xspace}

\def\Lbbar        {{\ensuremath{\Lbar{}^0_\bquark}}\xspace}

\def\BF         {{\ensuremath{\mathcal{B}}}\xspace}

\newcommand{\decay}[2]{\ensuremath{#1\!\to #2}\xspace} 

\def\to                 {\ensuremath{\rightarrow}\xspace}

\def\bsll     {\decay{\bquark}{\squark \ell^+ \ell^-}}

\def\AT#1     {\ensuremath{A_{\mathrm{T}}^{#1}}\xspace}           

\def\C#1      {\ensuremath{\mathcal{C}_{#1}}\xspace}                       
\def\Cp#1     {\ensuremath{\mathcal{C}_{#1}^{'}}\xspace}                    
\def\Ceff#1   {\ensuremath{\mathcal{C}_{#1}^{\mathrm{(eff)}}}\xspace}        
\def\Cpeff#1  {\ensuremath{\mathcal{C}_{#1}^{'\mathrm{(eff)}}}\xspace}       
\def\Ope#1    {\ensuremath{\mathcal{O}_{#1}}\xspace}                       
\def\Opep#1   {\ensuremath{\mathcal{O}_{#1}^{'}}\xspace}                    

\newcommand{\aunit}[1]{\ensuremath{\text{\,#1}}}       

\newcommand{\tev}{\aunit{Te\kern -0.1em V}\xspace}
\newcommand{\gev}{\aunit{Ge\kern -0.1em V}\xspace}
\newcommand{\mev}{\aunit{Me\kern -0.1em V}\xspace}
\newcommand{\kev}{\aunit{ke\kern -0.1em V}\xspace}
\newcommand{\ev}{\aunit{e\kern -0.1em V}\xspace}
 
\newcommand{\mevc}{\ensuremath{\aunit{Me\kern -0.1em V\!/}c}\xspace}
\newcommand{\gevc}{\ensuremath{\aunit{Ge\kern -0.1em V\!/}c}\xspace}
\newcommand{\mevcc}{\ensuremath{\aunit{Me\kern -0.1em V\!/}c^2}\xspace}
\newcommand{\gevcc}{\ensuremath{\aunit{Ge\kern -0.1em V\!/}c^2}\xspace}

\def\fb   {\ensuremath{\aunit{fb}}\xspace}
\def\invfb   {\ensuremath{\fb^{-1}}\xspace}

\def\gsim{{~\raise.15em\hbox{$>$}\kern-.85em
          \lower.35em\hbox{$\sim$}~}\xspace}
\def\lsim{{~\raise.15em\hbox{$<$}\kern-.85em
          \lower.35em\hbox{$\sim$}~}\xspace}

\def\sPlot{\mbox{\em sPlot}\xspace}

\def\evtgen     {\mbox{\textsc{EvtGen}}\xspace}

\def\geant      {\mbox{\textsc{Geant4}}\xspace}

\def\photos     {\mbox{\textsc{Photos}}\xspace}

\def\pythia     {\mbox{\textsc{Pythia}}\xspace}

\def\tell1  {TELL1\xspace}
\def\ukl1   {UKL1\xspace}

\usepackage{cite} 
\usepackage{mciteplus}

\newcommand{\sigdecay}{\decay{\Bz}{\Dz \Dzb\Kp\pim}}
\newcommand{\condecay}{\decay{\Bz}{\Dstarm \Dz \Kp}}
\usepackage{subcaption}
\usepackage{wrapfig}

\begin{document}

\renewcommand{\thefootnote}{\fnsymbol{footnote}}
\setcounter{footnote}{1}


\begin{titlepage}
\pagenumbering{roman}

\vspace*{-1.5cm}
\centerline{\large EUROPEAN ORGANIZATION FOR NUCLEAR RESEARCH (CERN)}
\vspace*{1.5cm}
\noindent
\begin{tabular*}{\linewidth}{lc@{\extracolsep{\fill}}r@{\extracolsep{0pt}}}
\ifthenelse{\boolean{pdflatex}}
{\vspace*{-1.5cm}\mbox{\!\!\!\includegraphics[width=.14\textwidth]{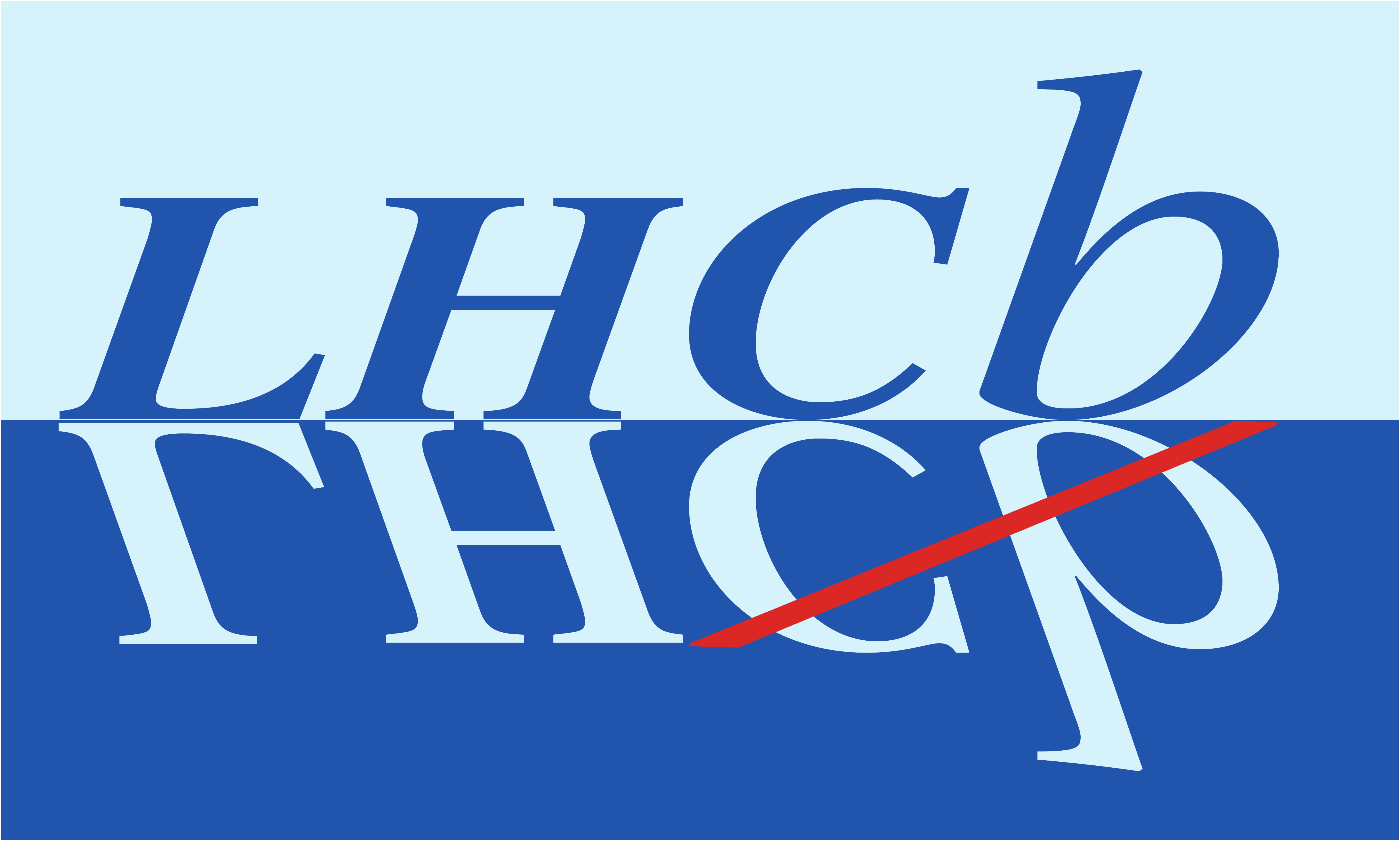}} & &}%
{\vspace*{-1.2cm}\mbox{\!\!\!\includegraphics[width=.12\textwidth]{figs/lhcb-logo.eps}} & &}%
\\
 & & CERN-EP-2020-112 \\  
 & & LHCb-PAPER-2020-015 \\  
 & & 9 July 2020 \\ 
 & & \\
\end{tabular*}

\vspace*{4.0cm}

{\normalfont\bfseries\boldmath\huge
\begin{center}
  \papertitle 
\end{center}
}

\vspace*{2.0cm}

\begin{center}
\paperauthors\footnote{Authors are listed at the end of this paper.}
\end{center}

\vspace{\fill}

\begin{abstract}
  \noindent
The first observation of the decay \sigdecay
is reported using proton-proton collision data corresponding to an integrated luminosity of 4.7 \invfb
collected by the \lhcb experiment in 2011, 2012 and 2016. The measurement is performed in the full kinematically allowed range of the decay outside of the \Dstarm region.
The ratio of the branching fraction relative to that of the control channel \condecay is measured to be 
\mbox{$\mathcal{R} = (14.2 \pm 1.1 \pm 1.0)\%$}, where the first uncertainty is statistical and the second is systematic. The absolute branching fraction of \sigdecay decays is thus determined to be 
\mbox{$\BF(\sigdecay) = (3.50 \pm 0.27 \pm 0.26 \pm 0.30) \times 10^{-4}$}, where the third uncertainty is due to the branching fraction of the control channel.
This decay mode is expected to provide insights to spectroscopy and the charm-loop contributions in rare semileptonic decays.
  
\end{abstract}

\vspace*{2.0cm}

\begin{center}
  Published in 
  Phys.~Rev.~D \textbf{102}, 051102(R) (2020) 
\end{center}

\vspace{\fill}

{\footnotesize 
\centerline{\copyright~\papercopyright. \href{\paperlicenceurl}{\paperlicence}.}}
\vspace*{2mm}

\end{titlepage}


\newpage
\setcounter{page}{2}
\mbox{~}
%
%
%
%


\renewcommand{\thefootnote}{\arabic{footnote}}
\setcounter{footnote}{0}

\cleardoublepage

\newcounter{zzz}
\pagestyle{plain} 
\setcounter{page}{1}
\pagenumbering{arabic}

\begin{figure}[!b]
    \centering
    \includegraphics[scale=1.2]{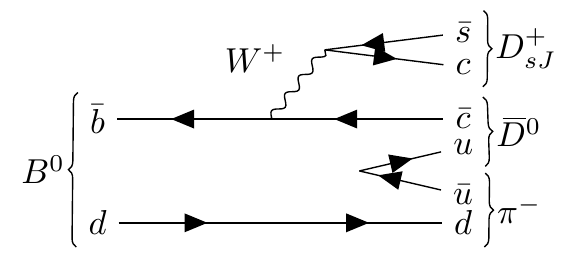}
    \includegraphics[scale=1.2]{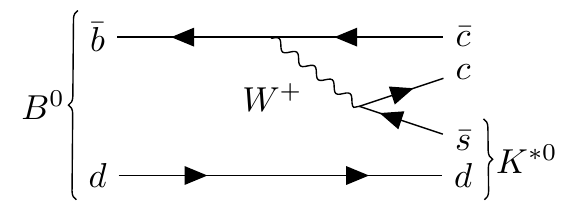}
    \caption{\small Feynman diagrams of the external (left) and internal (right) W emission contributing to \sigdecay decays.}
    \label{fig:feynman}
\end{figure}

\noindent The family of $\decay{B}{\D^{(\ast)} \Db{}^{(\ast)} K}$ and $\decay{B}{\D^{(\ast)} \Db{}^{(\ast)} K\pi}$ decays, each with two charm hadrons and a kaon in the final state, proceed at quark level through  Cabibbo-Kobayashi-Maskawa favoured $b\to c\overline{c} s$ transitions. These transitions occur with either an external or internal $W$ emission process, as shown in Fig.~\ref{fig:feynman}, offering the opportunity to search for new $c\overline{s}$ or $c\overline{c}$ states. In addition, measurements of the amplitude structure of the $\D^{(\ast)}\Db{}^{(\ast)}$ system in these processes can provide important information to calculations of the $c\bar{c}$ contribution above the open-charm threshold in \bsll decays~\cite{Khodjamirian:2010vf}. There is considerable debate whether the theoretical uncertainties associated with these long-distance contributions~\cite{Lyon:2014hpa,Ciuchini:2015qxb,Bobeth:2008ij,Blake:2017fyh} could alleviate the tensions in a wide range of measurements involving \bsll transitions~\cite{LHCb-PAPER-2020-002, Sirunyan:2017dhj, Aaboud:2018krd, Wehle:2016yoi, Aubert:2006vb, Aaltonen:2011ja, LHCb-PAPER-2018-029, LHCb-PAPER-2015-023, LHCb-PAPER-2016-012, LHCb-PAPER-2015-009, LHCb-PAPER-2014-006} with Standard Model predictions. Therefore, measurements that can provide input to these calculations are of the utmost importance. 

Although measurements involving $\decay{B}{\D^{(\ast)} \Db{}^{(\ast)} K}$ decays have been performed by the ALEPH, BaBar, Belle and LHCb collaborations~\cite{Barate:1998ch,Aubert:2003jq,Gokhroo:2006bt,Brodzicka:2007aa,delAmoSanchez:2010pg,LHCb-PAPER-2020-006}, no measurements involving \mbox{$\decay{B}{D^{(\ast)} \Db{}^{(\ast)} K\pi}$} transitions have been performed to date. The \sigdecay branching fraction, based on considerations of similar decay modes, is expected to be $\mathcal{O}(10^{-4})$, but the product of the branching fractions including the $\Dz \to \Km \pip$ charm meson decays is much smaller, at the level of $\mathcal{O}(10^{-7})$.

This paper presents the first observation of the \sigdecay decay, excluding contributions from \condecay transitions, with $\Dstarm\to \Dzb\pim$ decays.\footnote{The inclusion of charge-conjugate processes is implied throughout this paper unless otherwise noted.} The branching fraction of this decay is measured in the full kinematically allowed range of the decay outside of the \Dstarm region, relative to the control mode \condecay. After the decay of the \Dstarm meson via the strong interaction, signal and control modes present the same final-state particles $\Dz \Dzb \Kp \pim$. 
The measurement is performed using data collected with the LHCb detector in proton-proton collisions at centre-of-mass energies of $7\tev$ and $8\tev$ during 2011 and 2012 (Run~1), and $13\tev$ during 2016. The corresponding integrated luminosities for the years 2011, 2012 and 2016 are $1.0$, $2.0$ and $1.7\invfb$, respectively.

The \lhcb detector~\cite{LHCb-DP-2008-001,LHCb-DP-2014-002} is a
single-arm forward spectrometer covering the pseudorapidity range $2 < \eta < 5$, designed for
the study of particles containing \bquark\ or \cquark\ quarks. The detector elements that are particularly
relevant to this analysis are: a silicon-strip vertex detector surrounding the $pp$ interaction
region~\cite{LHCb-DP-2014-001} that allows \cquark\ and \bquark\ hadrons to be identified from their characteristically long
flight distance; a tracking system that provides a measurement of the momentum, $p$, of charged
particles~\cite{LHCb-DP-2013-003,LHCb-DP-2017-001}; and two ring-imaging Cherenkov detectors that are able to discriminate between
different species of charged hadrons~\cite{LHCb-DP-2012-003}. Photons, electrons and hadrons are identified by a calorimeter system consisting of scintillating-pad and preshower detectors, an electromagnetic and a hadronic calorimeter. Muons are identified by a system composed of alternating layers of iron and multiwire
proportional chambers. The online event selection is performed by a trigger, 
which consists of a hardware stage, based on information from the calorimeter and muon
systems, followed by a software stage, which applies a full event
reconstruction.
Events retained following the hardware trigger decision are split into two independent categories, those with a positive decision based on activity in the hadronic calorimeter associated with the signal candidate decay and those based on signatures from other particles in the event. The data are further split into two data taking categories for Run 1 and 2016 samples. 
The software trigger stage requires a two-, three- or four-track secondary vertex with a significant displacement from any primary $pp$ interaction vertex (PV).

Simulation is required to model the effects of the detector acceptance and the imposed selection requirements. It is also used to train multivariate classifiers for background suppression, and to obtain the shape of the invariant-mass distribution for candidate \Bz hadrons. In the simulation, $pp$ collisions are generated using \pythia~\cite{Sjostrand:2007gs} with a specific \lhcb configuration~\cite{LHCb-PROC-2010-056}. Decays of unstable particles are described by \evtgen~\cite{Lange:2001uf,LHCb-DP-2018-004}, in which final-state radiation is generated using \photos~\cite{Golonka:2005pn}. The interaction of the generated particles with the detector, and its response, are implemented using the \geant toolkit~\cite{Allison:2006ve, *Agostinelli:2002hh} as described in Ref.~\cite{LHCb-PROC-2011-006}.

The simulated samples of the signal- and control-mode decays are corrected to improve agreement with the data. A fit to the \Bz candidate invariant-mass distribution of the \mbox{\condecay} sample is performed using the \sPlot technique~\cite{splot-paper} to calculate weights that statistically remove background contributions. 
Subsequently, a correction to the simulation is derived as a function of event track multiplicity and impact parameter significance of the \Bz candidate with respect to the associated PV, by comparing \mbox{\condecay} candidates in simulation and background-subtracted data. 
In addition, the particle identification (PID) variables in the simulation are corrected using control data samples with the \textsc{Meerkat} software package~\cite{Meerkat,LHCb-DP-2018-001}. 

The $\Dz$ $(\Dzb)$ candidates are reconstructed in the $\Km\pip$ ($\Kp\pim$) final state, in a $\pm30\mevcc$ window around the known mass~\cite{PDG2019}. The $\Kp\pim$ candidates originating directly from the $\Bz$ decay are required to have an invariant mass below $1600\mevcc$ and are subsequently combined with the charm mesons to form the \Bz candidates. 

The selection comprises two stages. First, a loose selection is applied that relies on PID criteria to correctly identify charged kaons and pions, and on the flight distance significance of the $\Dz$ candidates to reject charmless backgrounds. The signal and control mode data samples are then split using the requirement \mbox{$|m(\Dzb \pim) - m(\Dzb) - [m_{0}(\Dstarm) - m_{0}(\Dzb)]| < (4 \times 0.724) \mevcc$}
to select candidates consistent with the \condecay hypothesis, where $m_0$ is the known mass of the particle~\cite{PDG2019} and $0.724\mevcc$ is the resolution of the \Dstarm contribution. 
To improve the mass resolution a global kinematic fit~\cite{Hulsbergen:2005pu} is performed constraining the mass of the $\Dz$ mesons to its known value. In this kinematic fit the $\Bz$ candidate is also constrained to originate from the associated PV. 

The second selection stage relies on two neural networks: one to identify good-quality $\Dz$ candidates from $\Bz$ meson decays (${\rm NN}_D$); and another to reduce the combinatorial background, which consists of candidates constructed from one or two random tracks in place of the \Kp and \pim from the \Bz meson decay (${\rm NN}_B$). A multilayer perceptron model is used, implemented using the \textsc{Keras} library~\cite{keras} in the \textsc{TensorFlow}~\cite{tensorflow} framework. These classifiers are trained separately for the Run~1 and 2016 data-taking periods and for each trigger category. 
The training and testing is performed
using the $k$-fold cross validation technique with $k = 10$ \cite{Blum:1999:BHB:307400.307439}. Simulated samples are used as a signal proxy and data from the sidebands of the \Dz or \Bz candidate invariant-mass distributions as the background proxy. Specifically, these are candidates outside of a $\pm 40\mevcc$ window around the known \Dz-meson mass~\cite{PDG2019} for the ${\rm NN}_D$ classifier and candidates satisfying $m(\Dz\Dzb K^+ \pi^-) > m_0(\Bz) + 100\mevcc$ for ${\rm NN}_B$.

The ${\rm NN}_D$ classifier is trained using 14 variables including PID information, kinematic properties and the decay topology of the tracks and $\Dz$ candidate.
Fourteen variables are also used to train the ${\rm NN}_B$ classifier, including the output of the two ${\rm NN}_D$ classifiers and other observables describing the topology and kinematics of the $\Bz$ meson decay.
As the ${\rm NN}_D$ classifier is an input to the ${\rm NN}_B$ classifier, a requirement is only placed on the output of the ${\rm NN}_B$ classifier. This threshold is optimised by maximising the figure of merit $\frac{N_S}{\sqrt{N_S+N_B}}$ separately in each of the two trigger categories and two data taking periods. Here $N_S$ is the expected signal yield calculated using the signal efficiency from the simulation and the estimated branching fraction based on branching fraction ratios of similar decays and the known branching fraction $\mathcal{B}(\Bp \to \Dz\Dzb\Kp)$~\cite{PDG2019}. The background yield $N_B$ is extrapolated from fits to the sidebands of the \Bz candidate invariant-mass distribution. The classifiers are found to be independent of the $m(\Dz\Dzb\Kp\pim)$ distribution.

The family of decays $H_b \to \Dz{}^{(*)}\Dzb{}^{(*)} H^{(*)}$, where $H_b$ is a beauty hadron and $H^{(*)}$ any one- or two-body collection of light or strange hadrons, is examined to search for possible background contributions. These are referred to as peaking backgrounds. Of these, four decay modes \mbox{$\Bp \to \Dz\Dzb\Kp$}, \mbox{$\Bp \to \Dstarz\Dzb\Kp$} (or equivalently, \mbox{$\Bp \to \Dz\Dstarzb\Kp$}), $\Bs \to \Dz\Dzb \phi$ and $\Lbbar \to \Dz\Dzb \bar{p} K^+$ are found to have substantial contributions to the signal channel. The \mbox{$\Bp \to \Dz\Dzb\Kp$} decays are removed using requirements on the three- and four-body invariant masses \mbox{$5220 < m(\Dz \Dzb \Kp) < 5340$ \mevcc} for candidates with $m(\Dz \Dzb \Kp \pim) > 5380 \mevcc$. The corresponding partially reconstructed decay $\Bp \to \Dstarz\Dzb\Kp$ is similarly removed with the requirement $5050 < m(\Dz \Dzb \Kp) < 5200$ \mevcc. Contributions from \mbox{$\Bs \to \Dz\Dzb \phi$} decays are suppressed using tighter PID requirements in the invariant-mass window \mbox{$5321 < m(\Dz \Dzb \Kp \Km) < 5411 \mevcc$}, where the \pim candidate is reconstructed under the \Km mass hypothesis. Similarly, \mbox{$\Lbbar \to \Dz\Dzb \bar{p} K^+$} candidates are removed using PID requirements for candidates satisfying \mbox{$5575 < m(\Dz \Dzb \Kp \bar{p}) < 5665 \mevcc$}, with the \pim candidate reconstructed using the \antiproton mass hypothesis. All of these backgrounds are reduced to negligible levels, and only the $\Bp \to \Dstarz\Dzb\Kp$ veto induces a sizeable signal loss with an efficiency of $93\%$.

A particularly challenging source of background is the modes $\Bz \to \Dz K^+\pi^- K^+\pi^-$, \mbox{$\Bz \to \Dzb K^-\pi^+ K^+\pi^-$} and \mbox{$\Bz \to K^-\pi^+ K^+\pi^- K^+\pi^-$}, so called single-charm and charmless backgrounds, respectively. Contributions from these decays are reduced by the flight distance criterion for the $\Dz$ mesons, but must be estimated carefully because they peak at the known $\Bz$ meson mass. 
The residual backgrounds are estimated from the sidebands of the \Dz invariant-mass distributions to be $10 \pm 7$ candidates. These candidates are subtracted from the yields during the fitting procedure described below.

The efficiency of the selections applied to the signal and control modes is calculated from simulated samples.
The selection efficiencies include the geometrical acceptance of the LHCb detector, the online trigger and event reconstruction, offline selections and the neural network classifiers.
For the signal mode, a single total efficiency is calculated and the resulting dependence on this efficiency model is considered as a systematic uncertainty. 
For the control mode, efficiency variations are seen over the phase space. Therefore, an efficiency is calculated for each candidate that depends on the two-dimensional Dalitz plot of the control mode decay.

Extended unbinned maximum-likelihood fits are performed to the \Bz candidate invariant-mass distributions of the signal and control channels in the range \mbox{$5235<m(\Dz\Dzb\Kp\pim)<5600\mevcc$}. The resolution of the $m(\Dz\Dzb\Kp\pim)$ distribution means that the contribution from partially reconstructed $B\to \Dstarz\Dzb\Kp\pim$ and $B\to \Dstarz\Dstarzb\Kp\pim$ decays is negligible in this fit range~\cite{suppl_mass}. 

The fit to the control mode is performed separately in the four data samples, corresponding to the two trigger categories and two data taking periods. The fit to the signal channel is performed simultaneously to these four categories.
The invariant-mass distributions for signal and control mode are modelled with a double-sided Crystal Ball function~\cite{Skwarnicki:1986xj}.
The parameters describing the tails of these distributions are fixed from fits to simulation separately for each of the four data samples. For the control mode, the mean and width of the mass distribution are determined directly from fits to the data subsamples. The resulting values are compared to those obtained on a fit to simulation to derive correction factors, which are subsequently used in the fits to the mass distribution of the signal channel. 
For the signal and control mode fits, the combinatorial background in each data sample is modelled with an exponential function with a slope allowed to vary in the fit. 
In the signal mode, the selections against the peaking backgrounds smoothly modify the shape of the mass distribution of the combinatorial background. This is accounted for by modulating the exponential function by an empirical correction from simulation. 
In the subsequent fits to the mass distribution of the signal candidates the ratio of branching fractions between the signal and control modes, $\mathcal{R}= \frac{\BF(\sigdecay)}{\BF(\condecay)}$, is expressed in terms of the signal yield in each of the four data samples as
\begin{equation}
    \mathcal{R} = \BF(\Dstarm \to \Dzb \pim)\times\left(\frac{N_{\rm sig} \varepsilon_{\rm con}}{N_{\rm con}\varepsilon_{\rm sig}}\right),
\end{equation}
where $N_{\rm sig}$ and $N_{\rm con}$ are the yields of the signal and control modes, respectively, and $\varepsilon_{\rm sig}$, $\varepsilon_{\rm con}$ are the corresponding efficiencies. The $\mathcal{R}$ parameter is determined from the simultaneous fit to the four data samples. The yield $N_{\rm con}$ and its uncertainty are propagated from the fit to the control mode with a Gaussian constraint. 

\begin{figure}[!tb]
\centering
    \begin{subfigure}[t]{.48\textwidth}
    \centering
    \includegraphics[width = 1.05\textwidth]{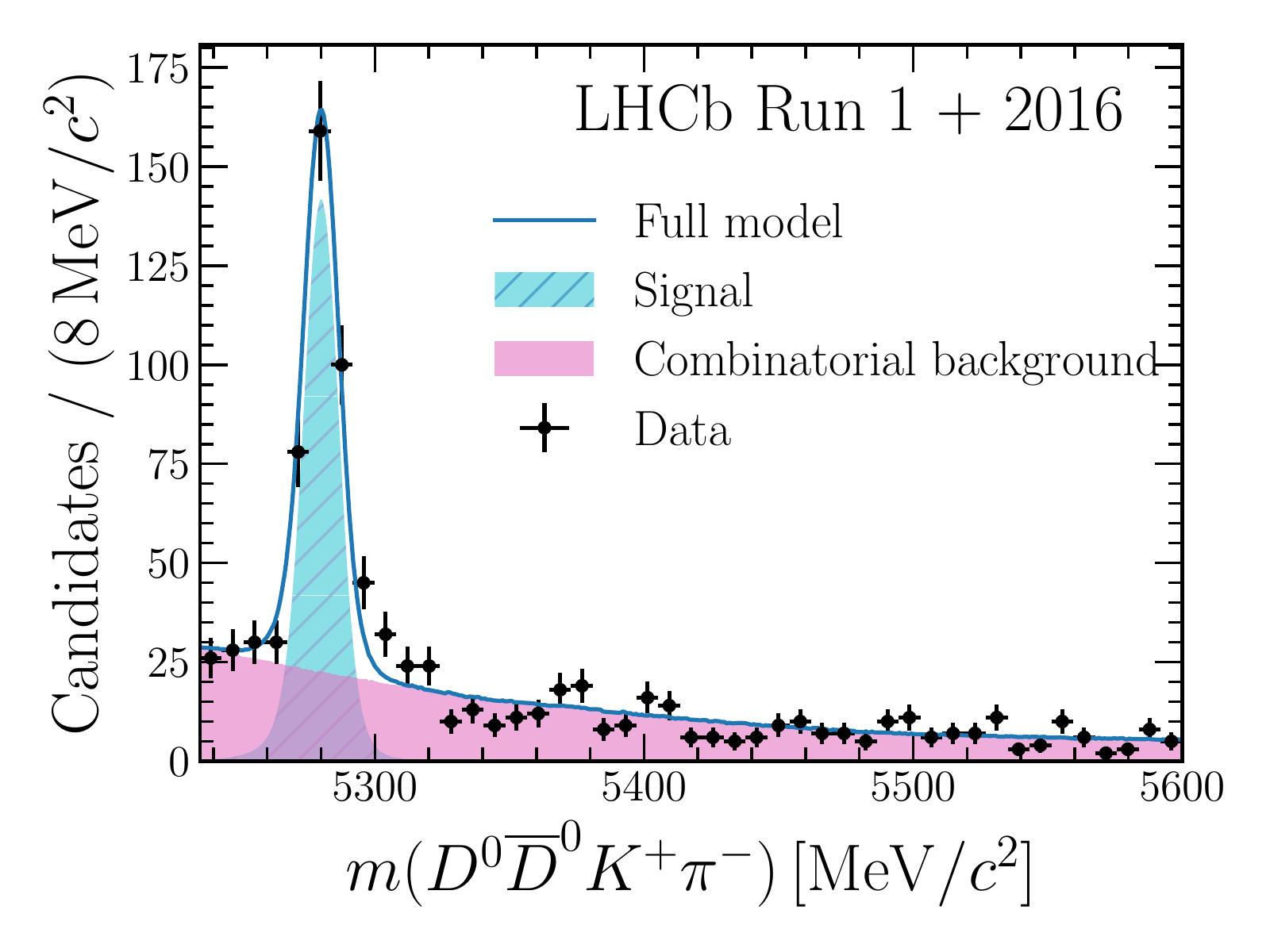}
    \end{subfigure}
    \begin{subfigure}[t]{.48\textwidth}
    \centering
    \includegraphics[width = 1.05\textwidth]{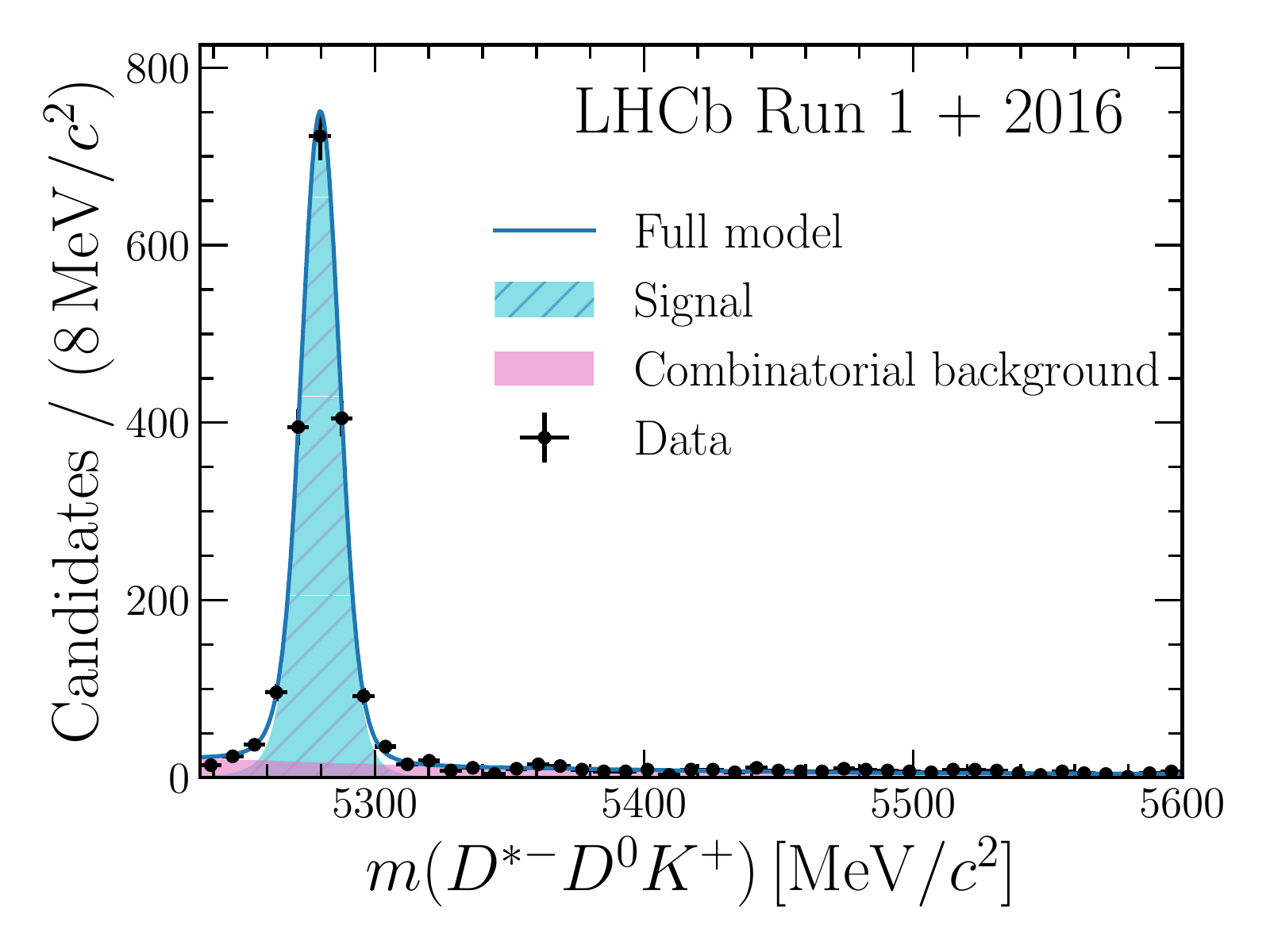}
    \end{subfigure}
\caption{\small Invariant-mass distributions and fit projections for $\Bz$ candidates in (left) the signal and (right) control mode for all subsamples combined. The data are shown as black points with error bars and the fit components are as described in the legends. The small single-charm and charmless background is included in the signal component.}
\label{fig:massfits}
\end{figure}

Invariant-mass distributions and fit projections of the \Bz candidates, summed over the trigger and data taking period subsamples, are shown in Fig.~\ref{fig:massfits}. In total $297 \pm 14$ signal and $1697 \pm 42$ control mode decays are found with a ratio of branching fractions $\mathcal{R} = (14.2 \pm 1.1)\%$, where the uncertainties are statistical only. 

\begin{figure}[!tb]
\hspace{-2em}
    \centering
    \begin{subfigure}[t]{.33\textwidth}
    \centering
    \includegraphics[width=1.05\textwidth,clip]{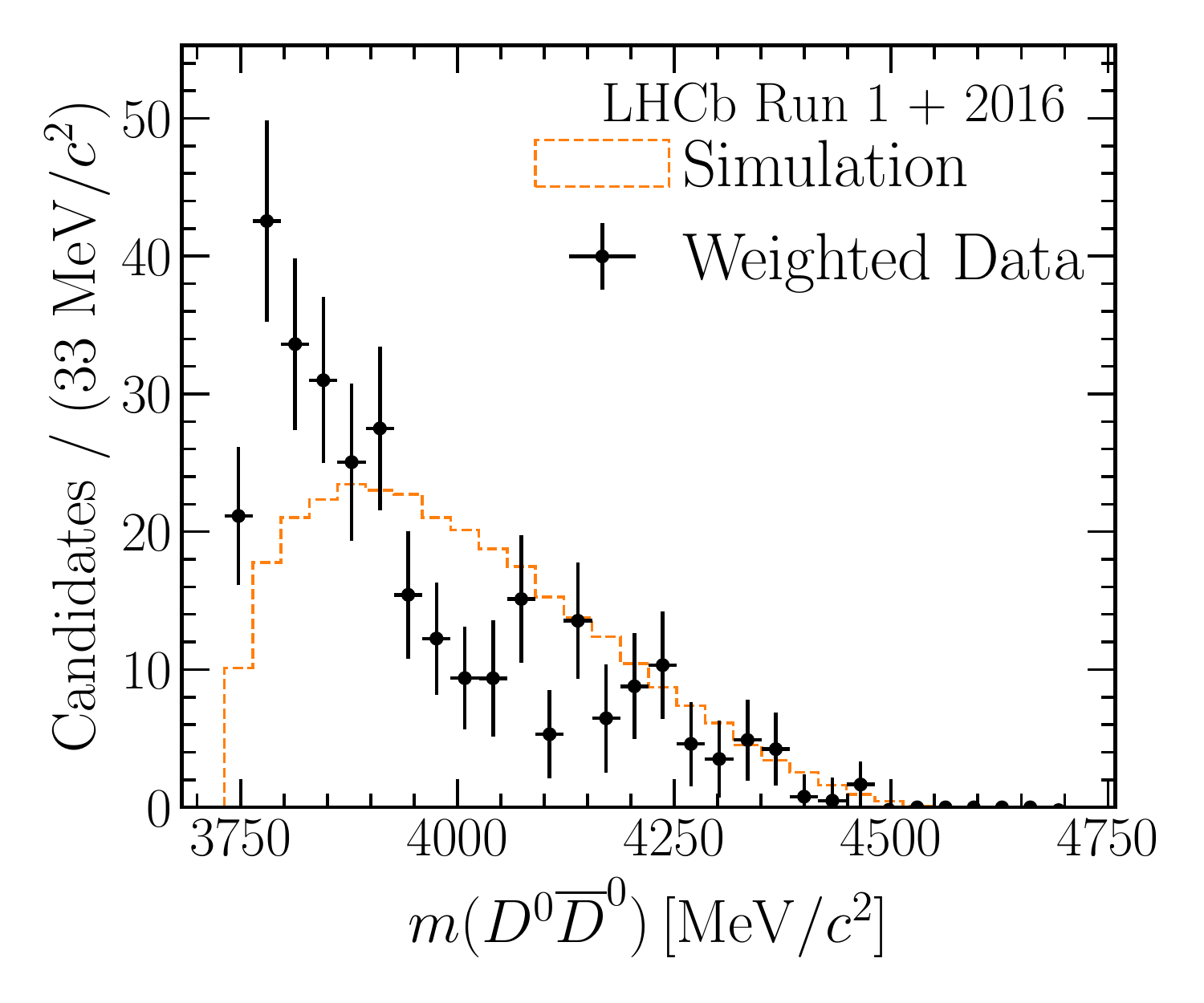}
    \end{subfigure}
    \begin{subfigure}[t]{.33\textwidth}
    \centering
    \includegraphics[width=1.05\textwidth,clip]{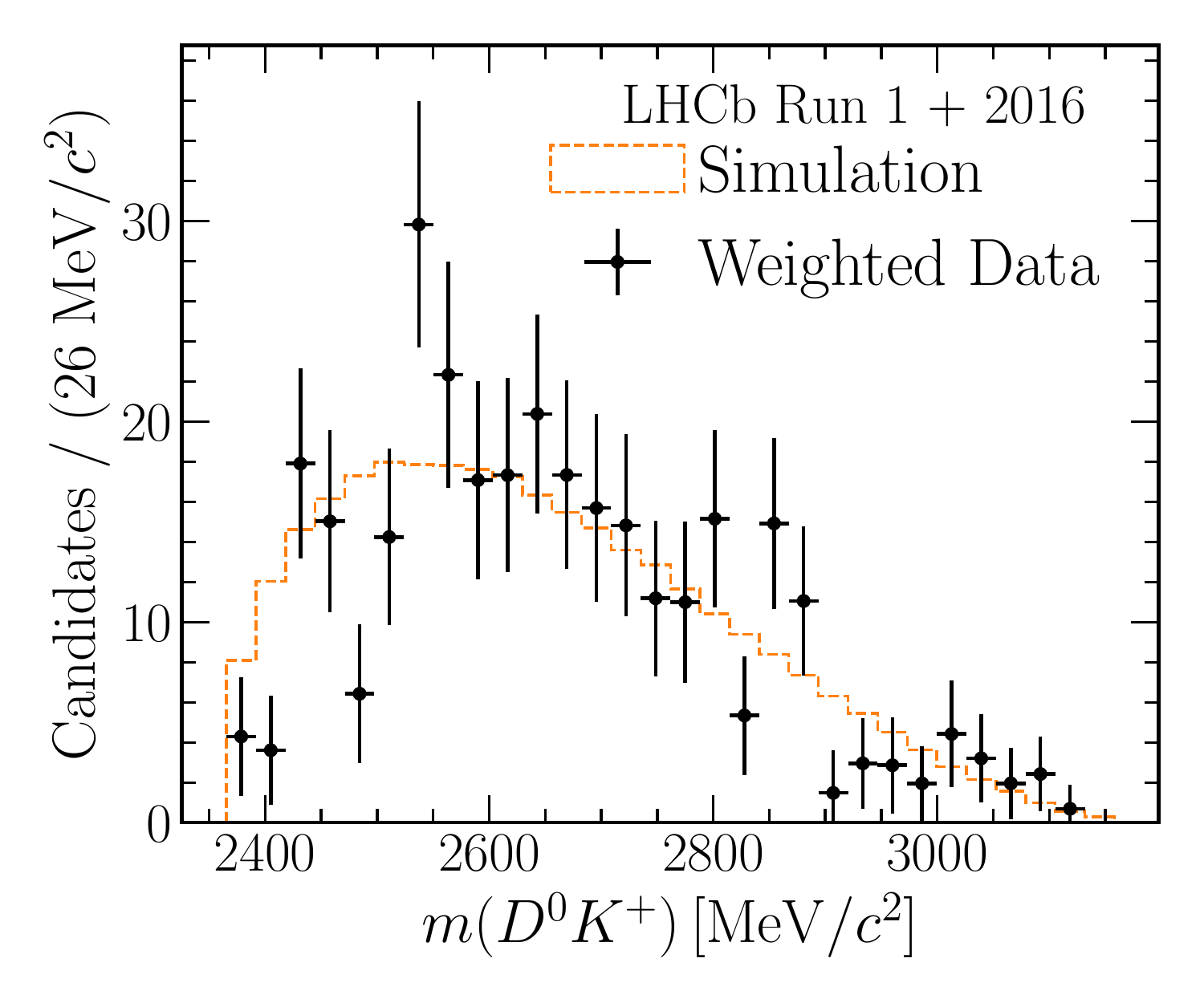}
    \end{subfigure}
    \begin{subfigure}[t]{.33\textwidth}
    \centering
    \includegraphics[width=1.05\textwidth,clip]{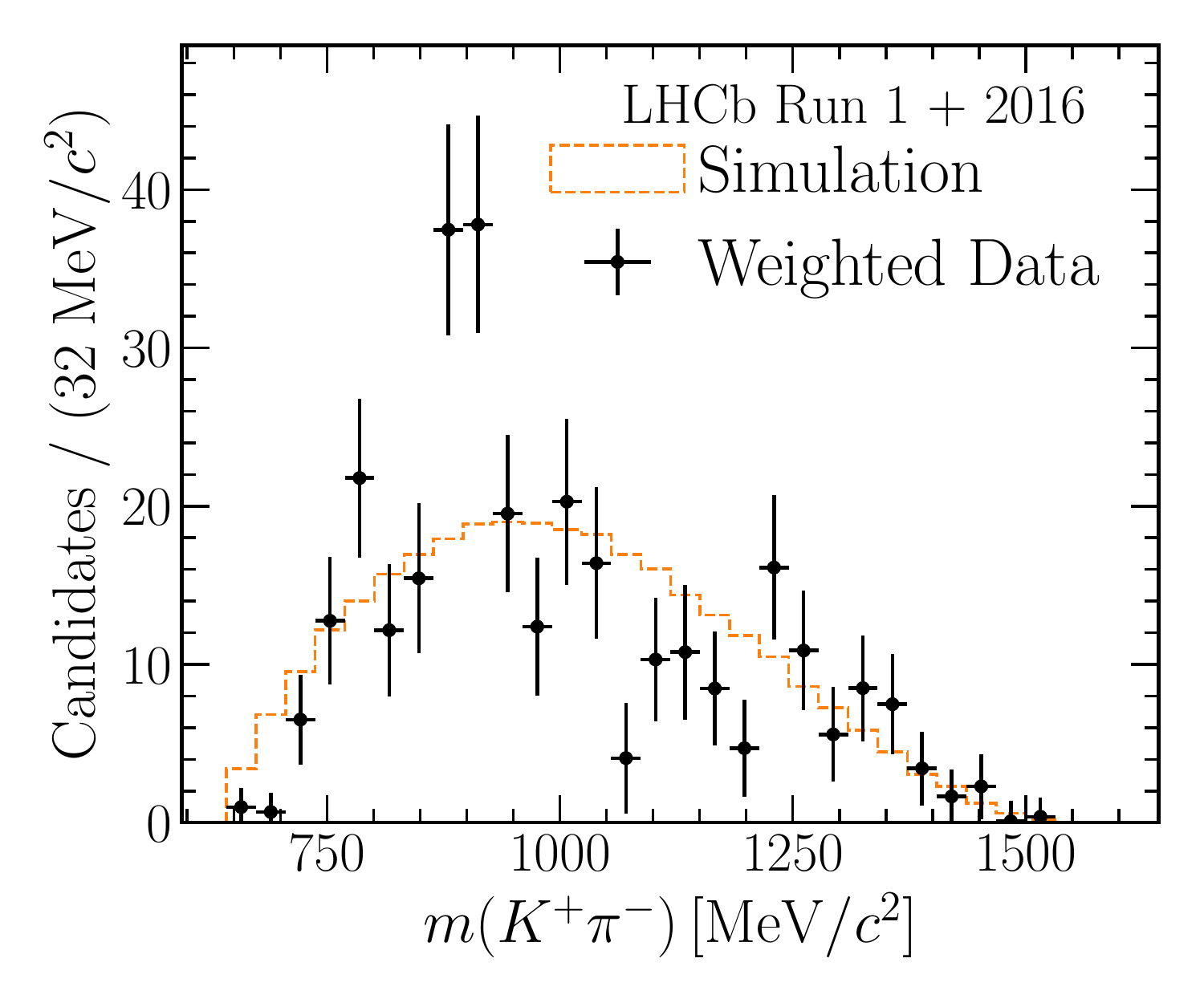}
    \end{subfigure}
    \caption{\small Projections of background-subtracted data (black points) in (left) $m(\Dz \Dzb)$, (centre) $m(\Dz\Kp)$ and (right) $m(\Kp \pim)$ with the phase-space only distribution (orange dashed line) superimposed for reference. The data contain a few single-charm and charmless background candidates.}
    \label{fig:proj}
\end{figure}
Figure~\ref{fig:proj} shows the background-subtracted~\cite{splot-paper} invariant-mass distributions of $m(\Dz \Dzb)$, $m(\Dz \Kp)$ and $m(\Kp \pim)$ overlaid with a simple phase-space distribution, including efficiency effects derived from simulation. There are hints of structures visible at the masses of the $\psi(3770)$, $D^*_{s2}(2573)^+$ and $D^*_{s(1,3)}(2860)^+$, and $K^*(892)^0$ states in the $m(\Dz \Dzb)$, $m(\Dz \Kp)$ and $m(\Kp \pim)$ distributions, respectively. Care should be taken with any interpretation of these projections because structures may be caused by reflections. Further analysis of these structures is left for future studies.

Several sources of systematic uncertainty are taken into account. The impact of using an averaged efficiency in the signal mode is considered by comparing the results using samples of $\decay{\Bz}{\Dz \Dzb \Kstarz}$ simulated events. An event-by-event correction to the efficiency is also considered, based on various three-dimensional parameterisations of the full five-dimensional phase space.
The fit model uncertainty is calculated by comparing the nominal background model to a polynomial form, and varying the signal shape parameters by sampling multivariate Gaussian distributions to account for the variance in the fit to simulation. The overall fit procedure is tested by generating pseudoexperiments from the nominal fit model using the measured values and fitting them with the same model. The results are compared to those from the nominal fit and no bias is observed.
The limited simulation sample size introduces a systematic uncertainty related to the spread in results obtained by varying the overall selection efficiencies within statistical uncertainties. Additionally, the weighting algorithm used to correct the simulation, as well as the data-driven method correcting the PID variables, introduce an associated statistical uncertainty. 
An uncertainty is also assigned to the estimation of single-charm and charmless background yields, by varying this contribution during the simultaneous fit to data. 
A correction is applied to the ${\rm NN}_B$ neural network classifier to account for possible mismodelling between data and simulation, and this uncertainty is calculated from the resulting difference in selection efficiencies. 
A small uncertainty is introduced due to the difference in the efficiency of selections applied to reconstruct candidates in signal and control modes. 
The systematic uncertainties are summarised in Table \ref{tab:systs}; they are summed in quadrature to give an overall relative systematic uncertainty on the ratio of branching fractions of $7.3\%$. 

\begin{table}[!tb]
\centering
\caption{\small Systematic uncertainties expressed as a percentage of the branching fraction ratio $\mathcal{R}$. The statistical uncertainty is included for comparison. The single-charm and charmless backgrounds are considered together.
}
\label{tab:systs}
\begin{tabular}{lc}
\hline
Source & Uncertainty (\%)\\
\hline
Signal model & 5.0 \\
Background model & 2.0 \\
Fixed fit parameters & 2.0 \\
Simulation sample size & 2.5 \\
Simulation weighting & 2.0 \\
PID weighting & 1.2 \\
Charmless backgrounds & 2.0 \\
Classifier modelling & 2.0 \\
Selection efficiency & 0.6 \\
\hline
Sum in quadrature & 7.3 \\
Statistical & 7.7 \\
\hline
\end{tabular}
\end{table}

In summary, the decay \sigdecay is observed for the first time, and its branching ratio relative to \condecay is measured to be
\begin{equation}
    \mathcal{R} = (14.2 \pm 1.1 \pm 1.0)\%,
\end{equation}
where the first uncertainty is statistical, and the second systematic. This measurement uses the full kinematically-allowed range of \sigdecay outside of the \Dstarm region, including the entire $\Kp \pim$ mass range, encompassing the $\Kstar(892)^0$ resonance and the broad $
\Kp \pim$ $S$-wave. The most precise measurement of the branching fraction of \condecay decays, performed by the BaBar collaboration, is  \mbox{$\mathcal{B}(\condecay) = (2.47 \pm 0.21) \times 10^{-3}$~\cite{delAmoSanchez:2010pg}}. Substituting in this value gives
\begin{equation}
    \mathcal{B}(\sigdecay) = (3.50 \pm 0.27 \pm 0.26 \pm 0.30) \times 10^{-4},
\end{equation}
where the third uncertainty comes from the uncertainty on the branching fraction \mbox{$\mathcal{B}(\condecay)$}.
Recently, the LHCb collaboration performed a measurement of the ratio of branching fractions $\frac{\BF(\condecay)}{\BF(\Bz\to \Dz\Dm\Kp)}$~\cite{LHCb-PAPER-2020-006}. However, the current precision on the branching fraction of the decay $\Bz\to \Dz\Dm\Kp$~\cite{PDG2019} does not yet allow for a more precise measurement of the decay rate $\BF(\condecay)$.
The results in this paper provide a crucial first step towards studying the rich resonant structure of these decays. An amplitude analysis will provide insights to both the spectroscopy of $c\overline{s}$ and $c\overline{c}$ states, and charm-loop contributions to \bsll decays.

\section*{Acknowledgements}
%
%
\noindent We express our gratitude to our colleagues in the CERN
accelerator departments for the excellent performance of the LHC. We
thank the technical and administrative staff at the LHCb
institutes.
We acknowledge support from CERN and from the national agencies:
CAPES, CNPq, FAPERJ and FINEP (Brazil); 
MOST and NSFC (China); 
CNRS/IN2P3 (France); 
BMBF, DFG and MPG (Germany); 
INFN (Italy); 
NWO (Netherlands); 
MNiSW and NCN (Poland); 
MEN/IFA (Romania); 
MSHE (Russia); 
MinECo (Spain); 
SNSF and SER (Switzerland); 
NASU (Ukraine); 
STFC (United Kingdom); 
DOE NP and NSF (USA).
We acknowledge the computing resources that are provided by CERN, IN2P3
(France), KIT and DESY (Germany), INFN (Italy), SURF (Netherlands),
PIC (Spain), GridPP (United Kingdom), RRCKI and Yandex
LLC (Russia), CSCS (Switzerland), IFIN-HH (Romania), CBPF (Brazil),
PL-GRID (Poland) and OSC (USA).
We are indebted to the communities behind the multiple open-source
software packages on which we depend.
Individual groups or members have received support from
AvH Foundation (Germany);
EPLANET, Marie Sk\l{}odowska-Curie Actions and ERC (European Union);
A*MIDEX, ANR, Labex P2IO and OCEVU, and R\'{e}gion Auvergne-Rh\^{o}ne-Alpes (France);
Key Research Program of Frontier Sciences of CAS, CAS PIFI, and the Thousand Talents Program (China);
RFBR, RSF and Yandex LLC (Russia);
GVA, XuntaGal and GENCAT (Spain);
the Royal Society
and the Leverhulme Trust (United Kingdom).

\setcounter{zzz}{\value{page}}



\renewcommand{\thetable}{S\arabic{table}}  

\renewcommand{\thefigure}{S\arabic{figure}}

\renewcommand{\theequation}{S\arabic{equation}}

\setcounter{figure}{0}

\setcounter{table}{0}

\setcounter{equation}{0}
\clearpage
\pagenumbering{roman}
\setcounter{section}{0}

{\noindent\bf\Large Supplemental Material} \\

\noindent This supplemental material includes additional information to that already provided in the main paper. Section~\ref{sec:supp_effcorr} details the correction applied to the selection efficiency of the control mode. Section~\ref{sec:supp_fits} shows the invariant-mass distributions and fit projections separately for the data subsamples. 

\section{Efficiency correction} \label{sec:supp_effcorr}
\begin{figure}[!htb]
\centering
    \begin{subfigure}[t]{.48\textwidth}
    \centering
    \includegraphics[width = 1.05\textwidth]{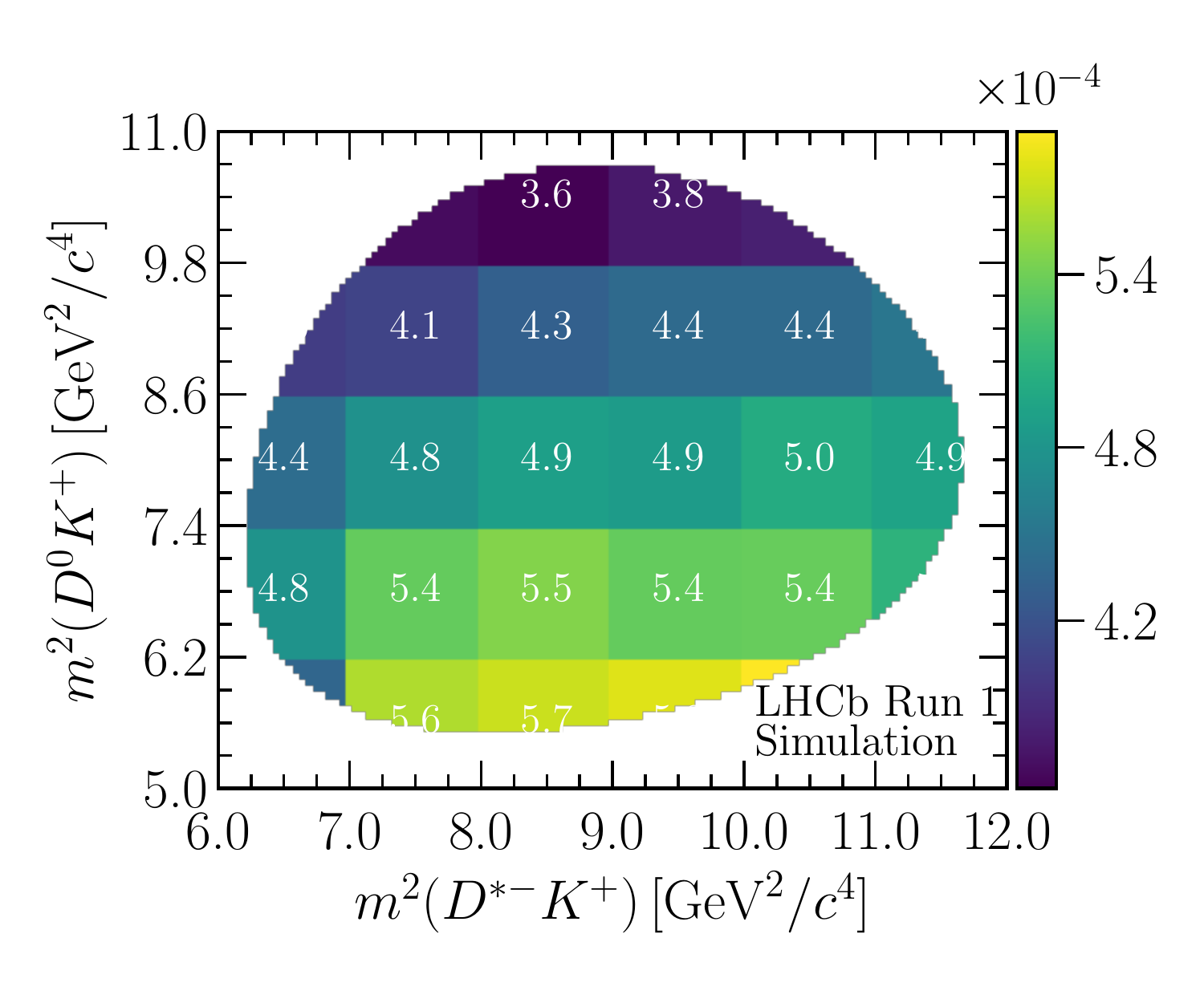}
    \end{subfigure}
    \begin{subfigure}[t]{.48\textwidth}
    \centering
    \includegraphics[width = 1.05\textwidth]{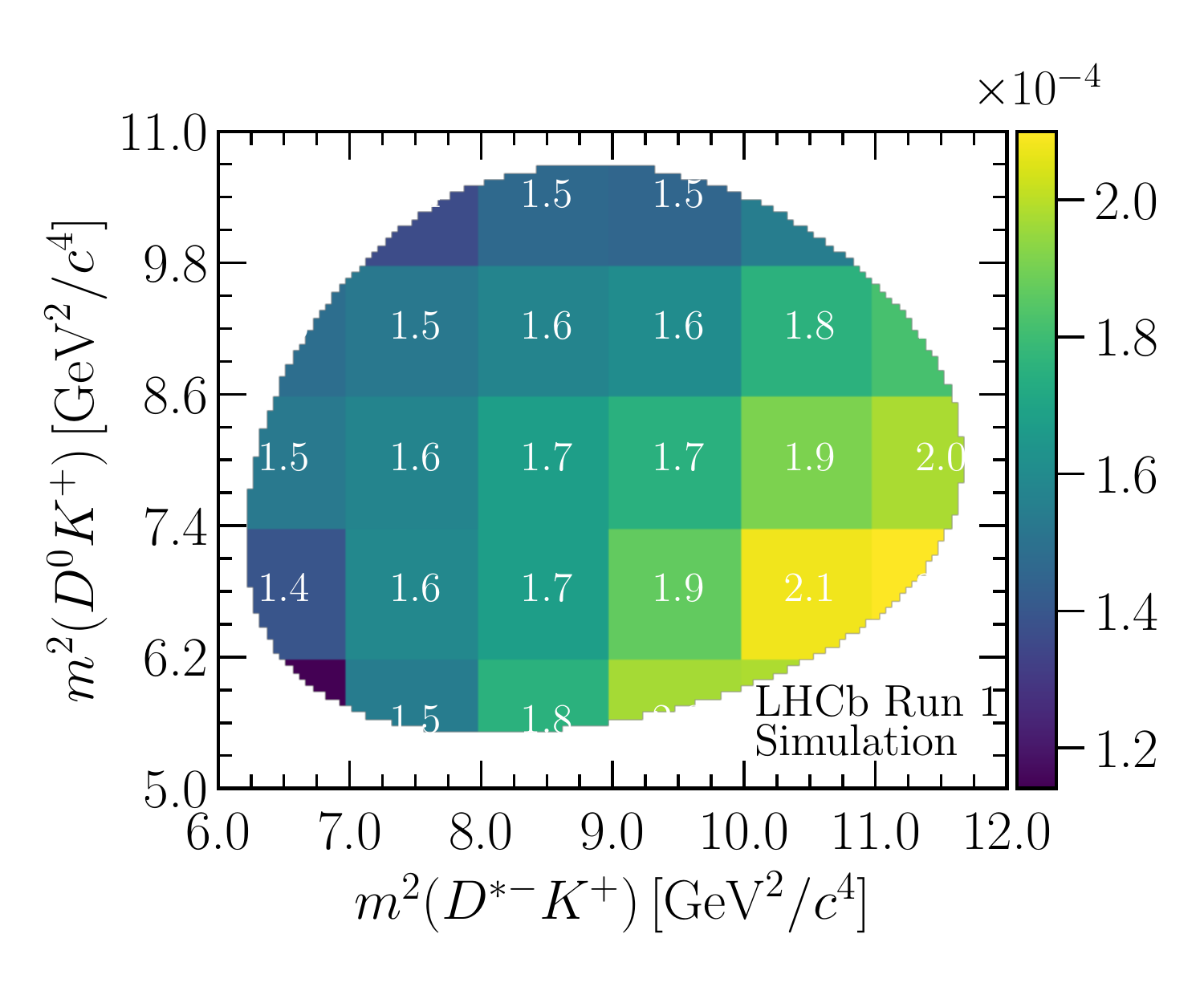}
    \end{subfigure}
    \begin{subfigure}[t]{.48\textwidth}
    \centering
    \includegraphics[width = 1.05\textwidth]{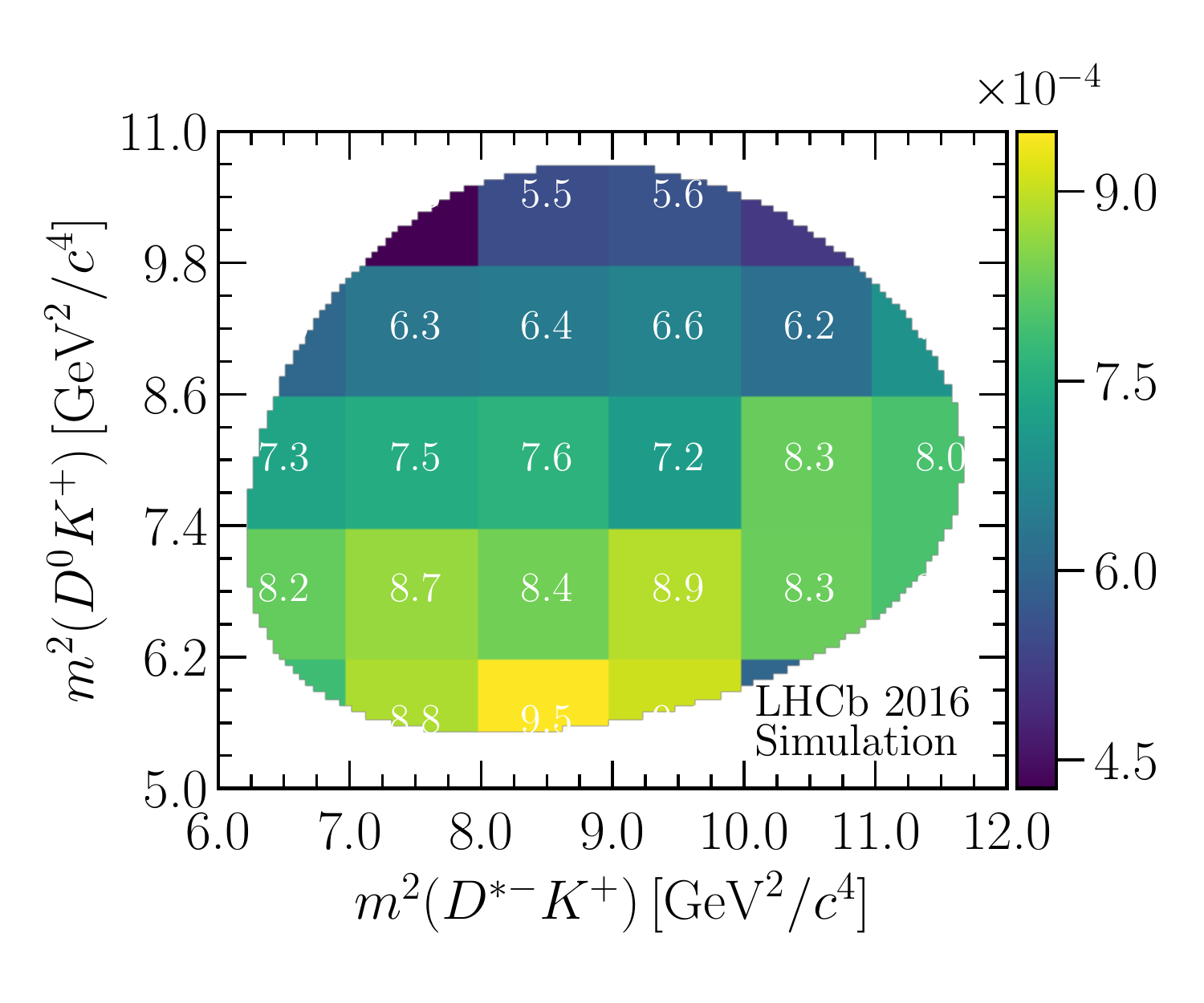}
    \end{subfigure}
    \begin{subfigure}[t]{.48\textwidth}
    \centering
    \includegraphics[width = 1.05\textwidth]{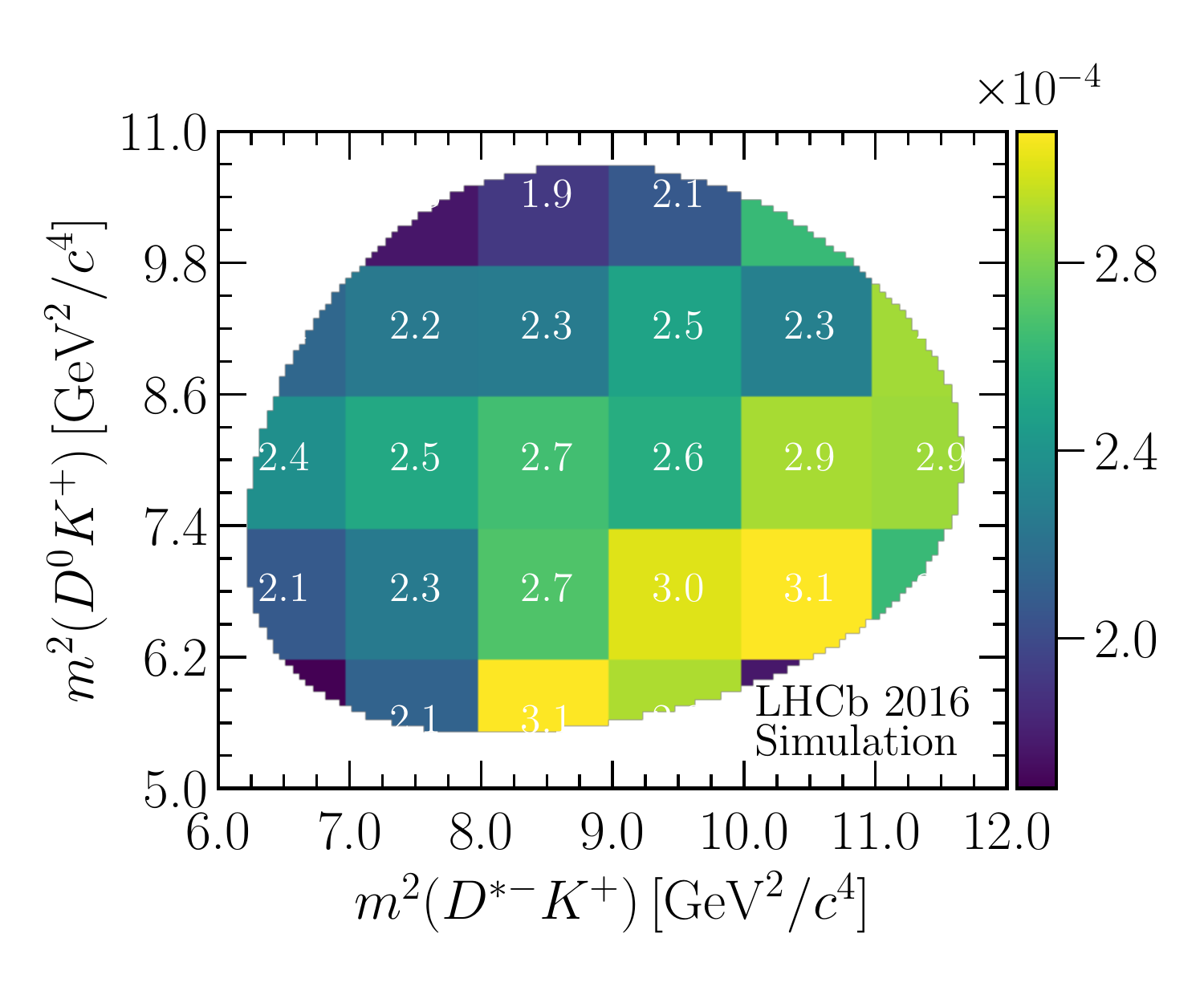}
    \end{subfigure}
  \caption{
    \small The efficiency $\varepsilon_{\rm{con}},$ as a function of position in the Dalitz plot of \condecay decays obtained from simulated samples. The top row shows Run 1 samples for (left) TOS and (right) TIS. The bottom row shows the same trigger categories for 2016 samples. }
  \label{fig:eff_corr}
\end{figure}

To account for variations across the Dalitz plot, \mbox{$m^2(\Dstarm \Kp)$ and $m^2(\Dz \Kp)$}, in the control mode, the efficiency is calculated, in each category, as
\begin{align}
    \varepsilon_{\rm con}(m^2(\Dstarm \Kp), m^2(\Dz \Kp)) = \frac{\sum_i w_i \varepsilon_i(m^2(\Dstarm \Kp), m^2(\Dz \Kp))}{\sum_i w_i}.
\end{align}
Here $w_i$ is the weight derived from the \sPlot technique, with the $\Bz$ candidate invariant mass distribution as the discriminating variable. The symbol $\varepsilon_i$ refers to the efficiency as a function of the Dalitz plot position for candidate $i$ in the control mode data sample. The efficiency variation across the Dalitz plot is displayed in Fig.~\ref{fig:eff_corr}. The average uncertainty across the bins of the Dalitz plane ranges from $1\rm{-}8\%$ for the four data samples. The two exclusive trigger categories are henceforth referred to as Trigger on Signal (TOS), for events triggered by the signal decay, and Trigger Independent of Signal (TIS), for events triggered by other particles in the event and that do not fall into the TOS category.

\newpage
\clearpage

\section{Mass fits} \label{sec:supp_fits}
\begin{figure}[!htb]
\centering
    \begin{subfigure}[t]{.48\textwidth}
    \centering
    \includegraphics[width = 1.05\textwidth]{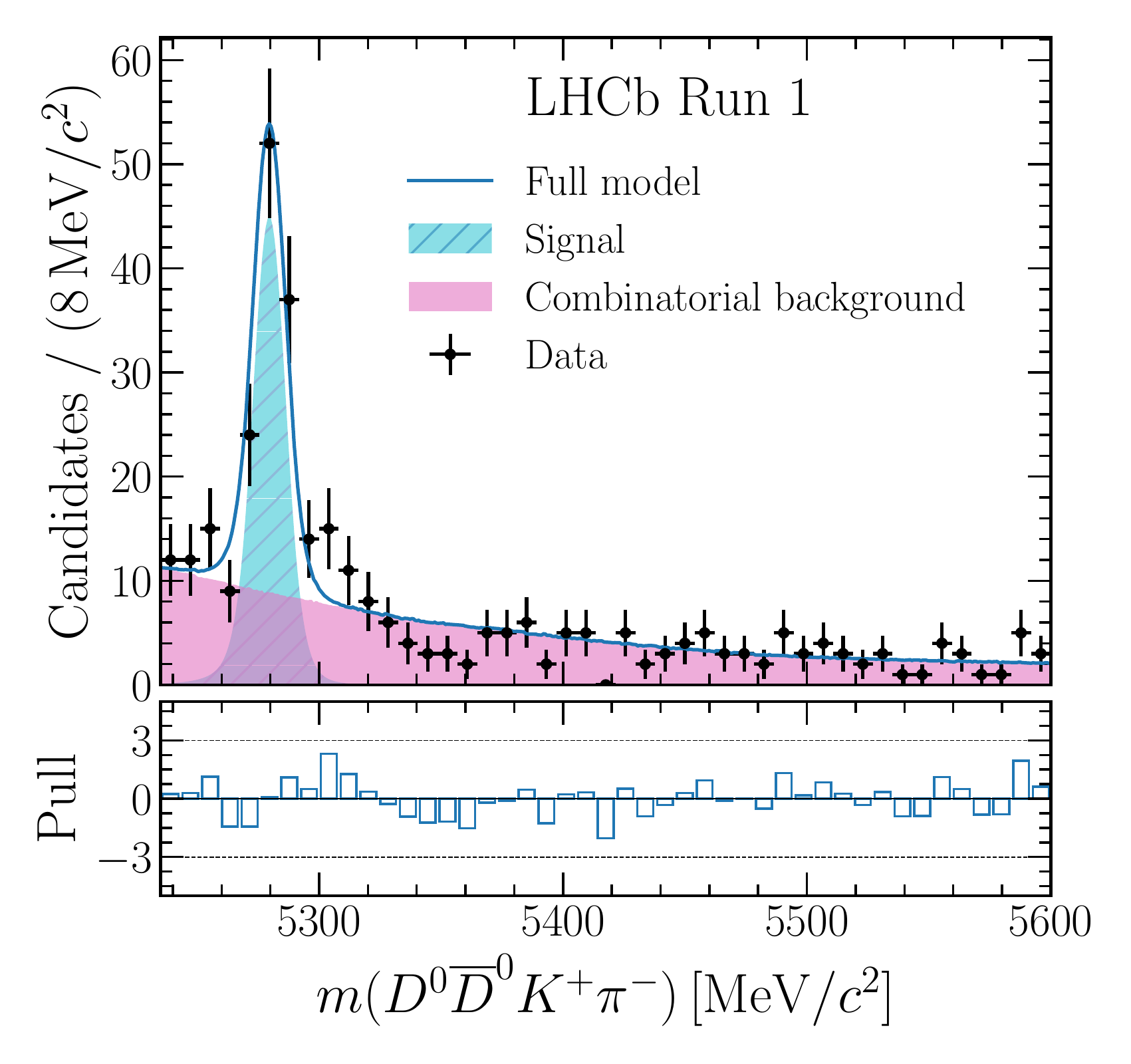}
    \end{subfigure}
    \begin{subfigure}[t]{.48\textwidth}
    \centering
    \includegraphics[width = 1.05\textwidth]{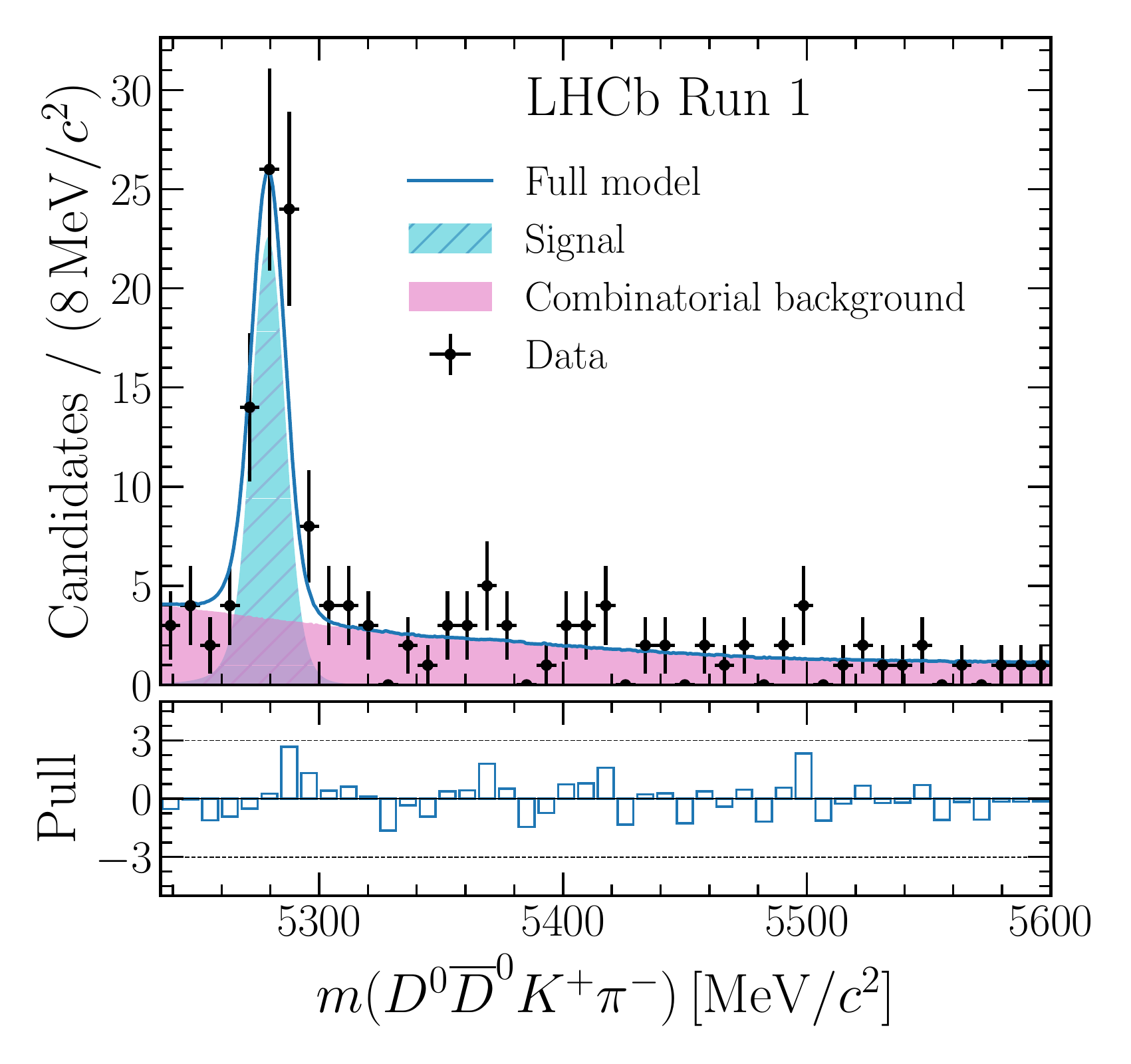}
    \end{subfigure}
    \begin{subfigure}[t]{.48\textwidth}
    \centering
    \includegraphics[width = 1.05\textwidth]{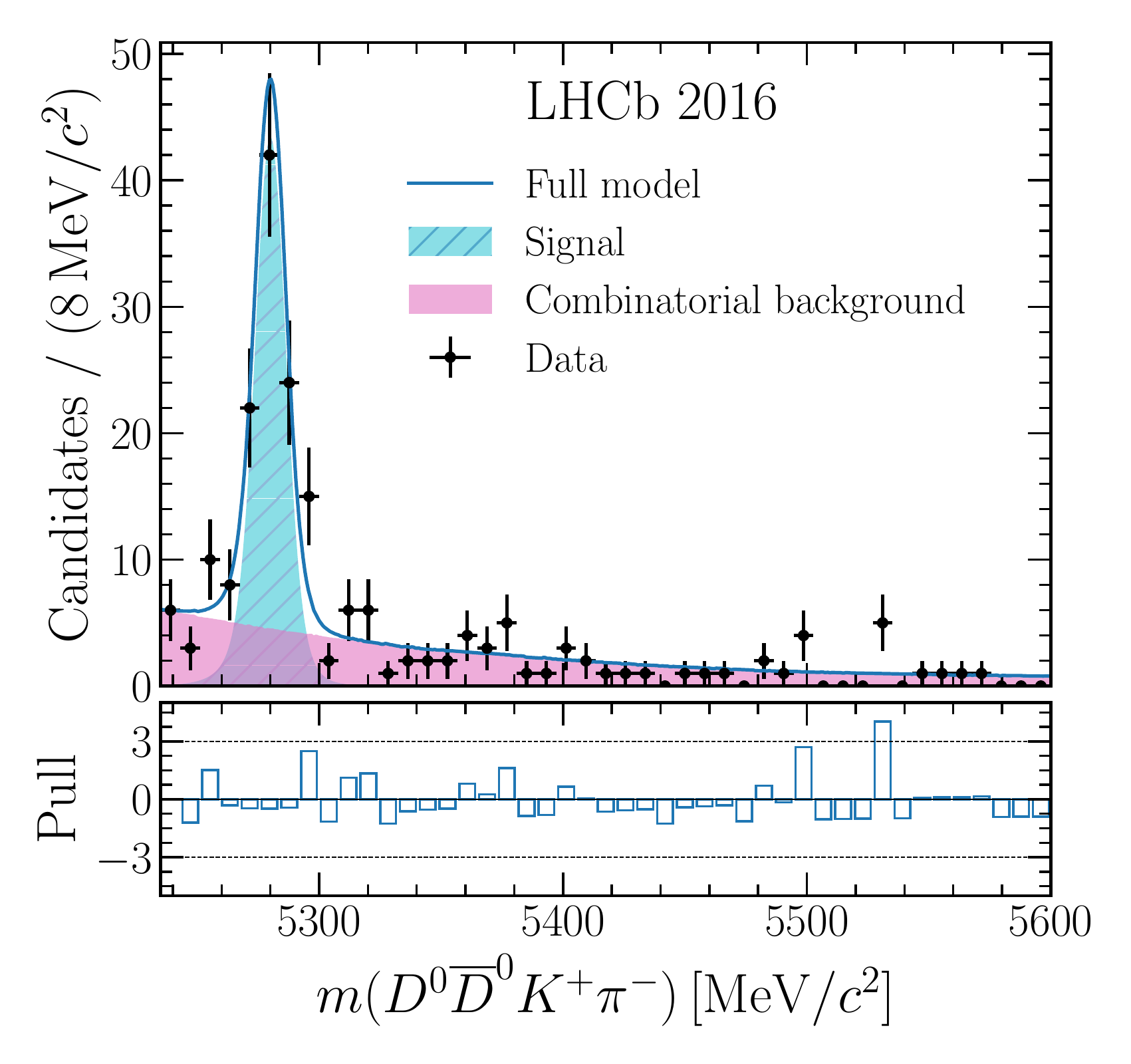}
    \end{subfigure}
    \begin{subfigure}[t]{.48\textwidth}
    \centering
    \includegraphics[width = 1.05\textwidth]{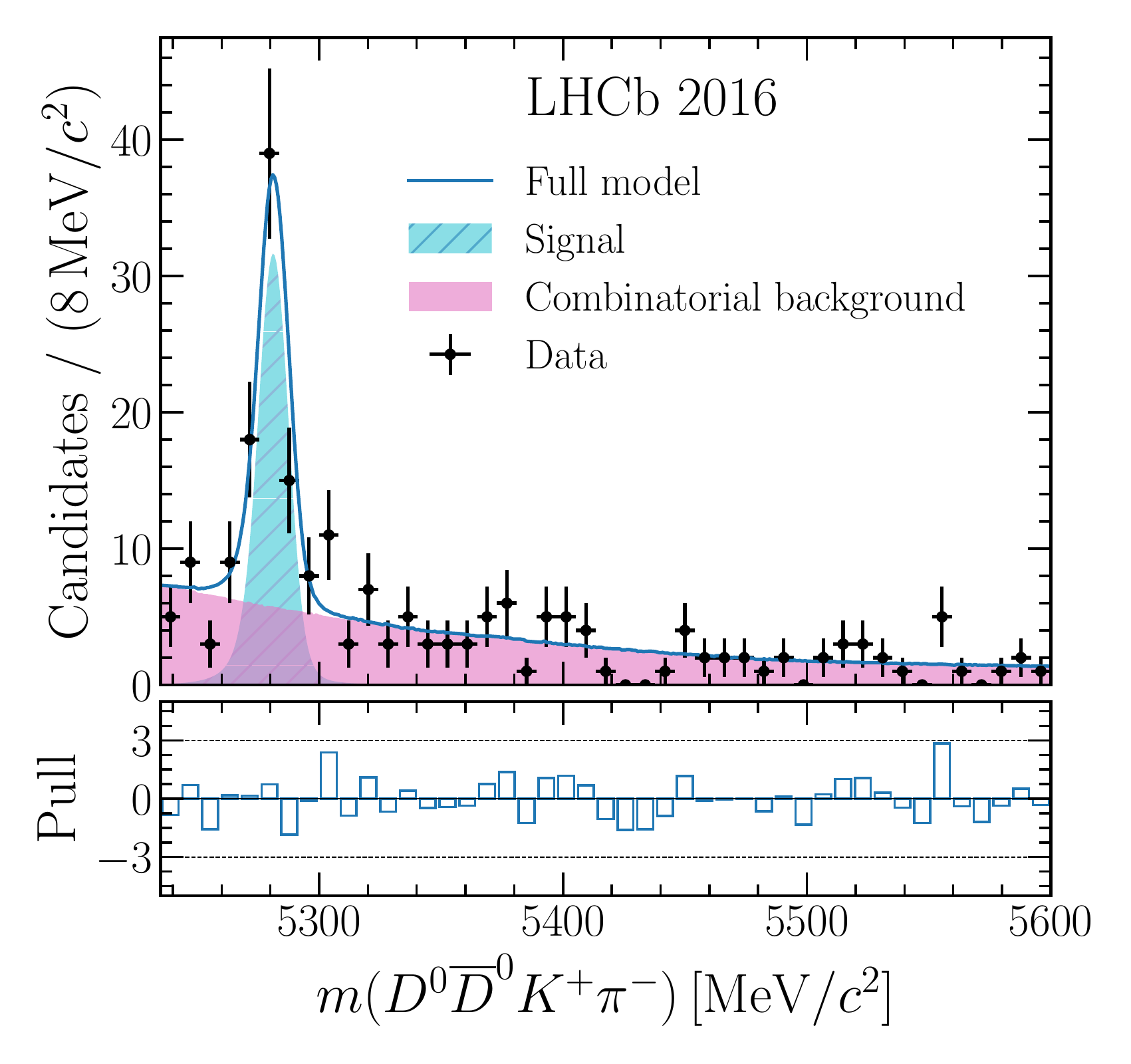}
    \end{subfigure}
  \caption{
    \small Invariant-mass distributions and fit projections for \sigdecay candidates, shown separately for each run period, (top) Run 1 and (bottom) 2016, and trigger category, (left) TOS and (right) TIS. The data are shown as black points with error bars and the fit components are as described in the legends. Pull projections are shown beneath each distribution. The data contain a few single-charm and charmless background candidates.}
  \label{fig:signal_fits}
\end{figure}

\begin{figure}[!htb]
\centering
    \begin{subfigure}[t]{.48\textwidth}
    \centering
    \includegraphics[width = 1.05\textwidth]{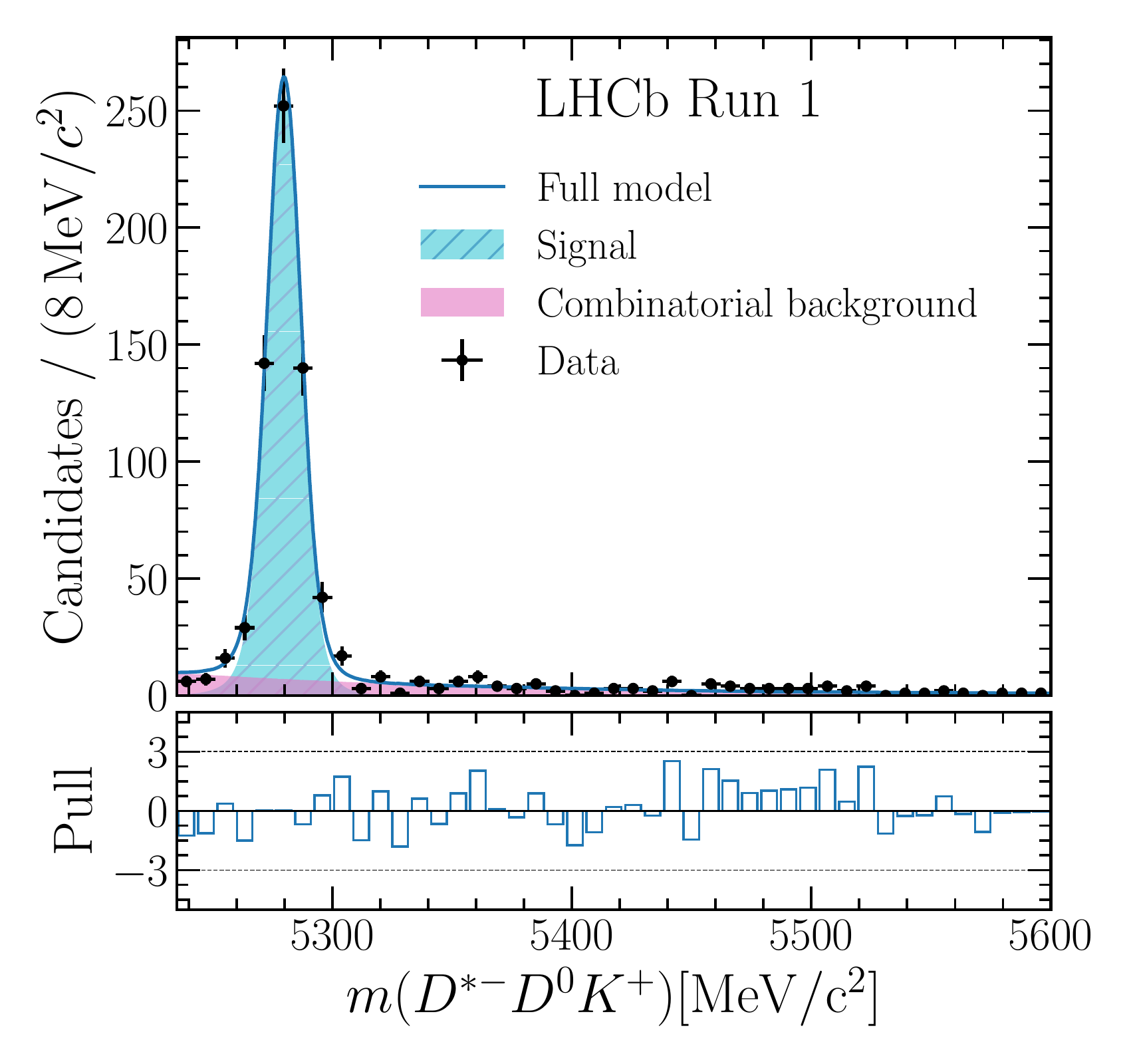}
    \end{subfigure}
    \begin{subfigure}[t]{.48\textwidth}
    \centering
    \includegraphics[width = 1.05\textwidth]{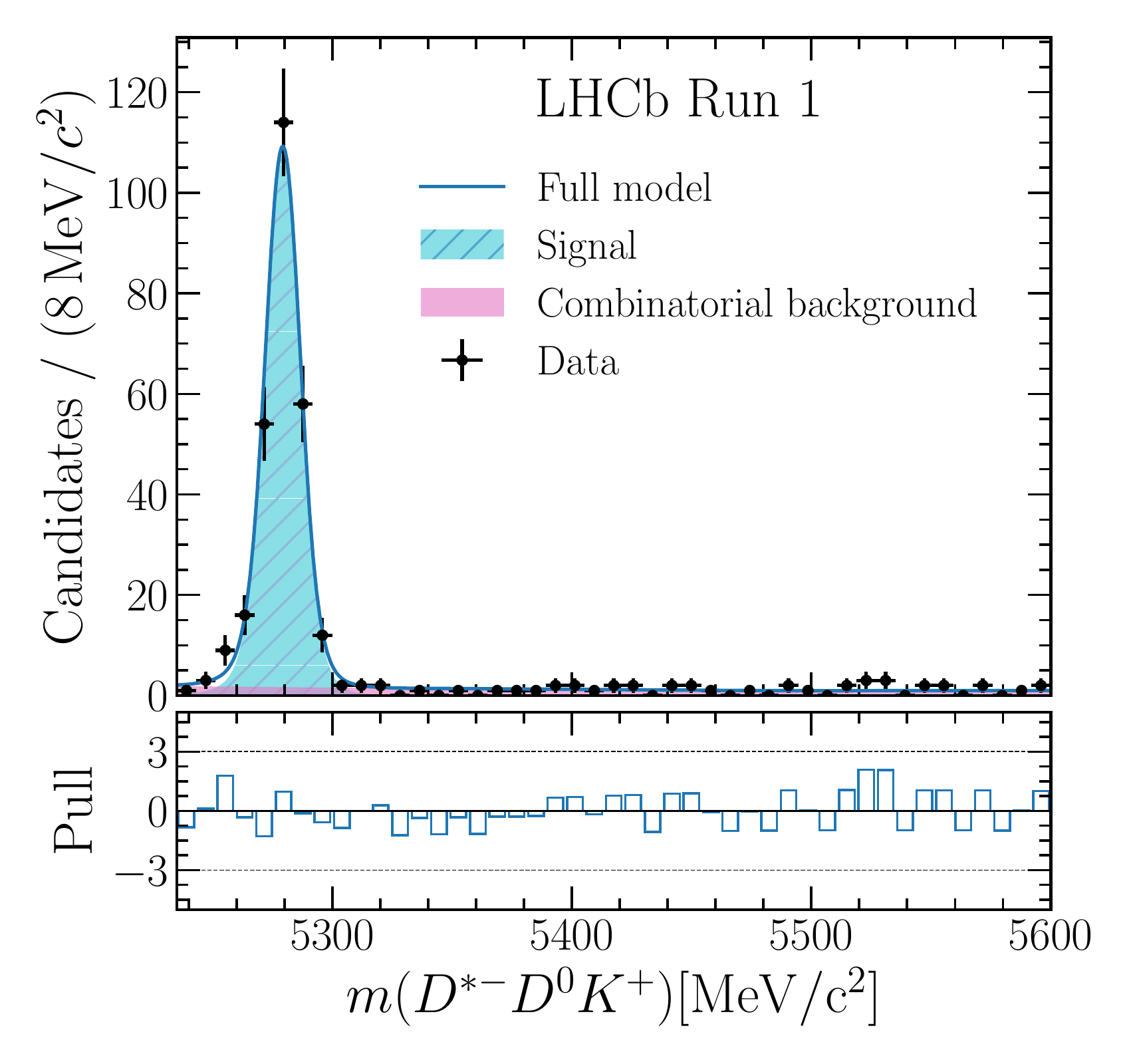}
    \end{subfigure}
    \begin{subfigure}[t]{.48\textwidth}
    \centering
    \includegraphics[width = 1.05\textwidth]{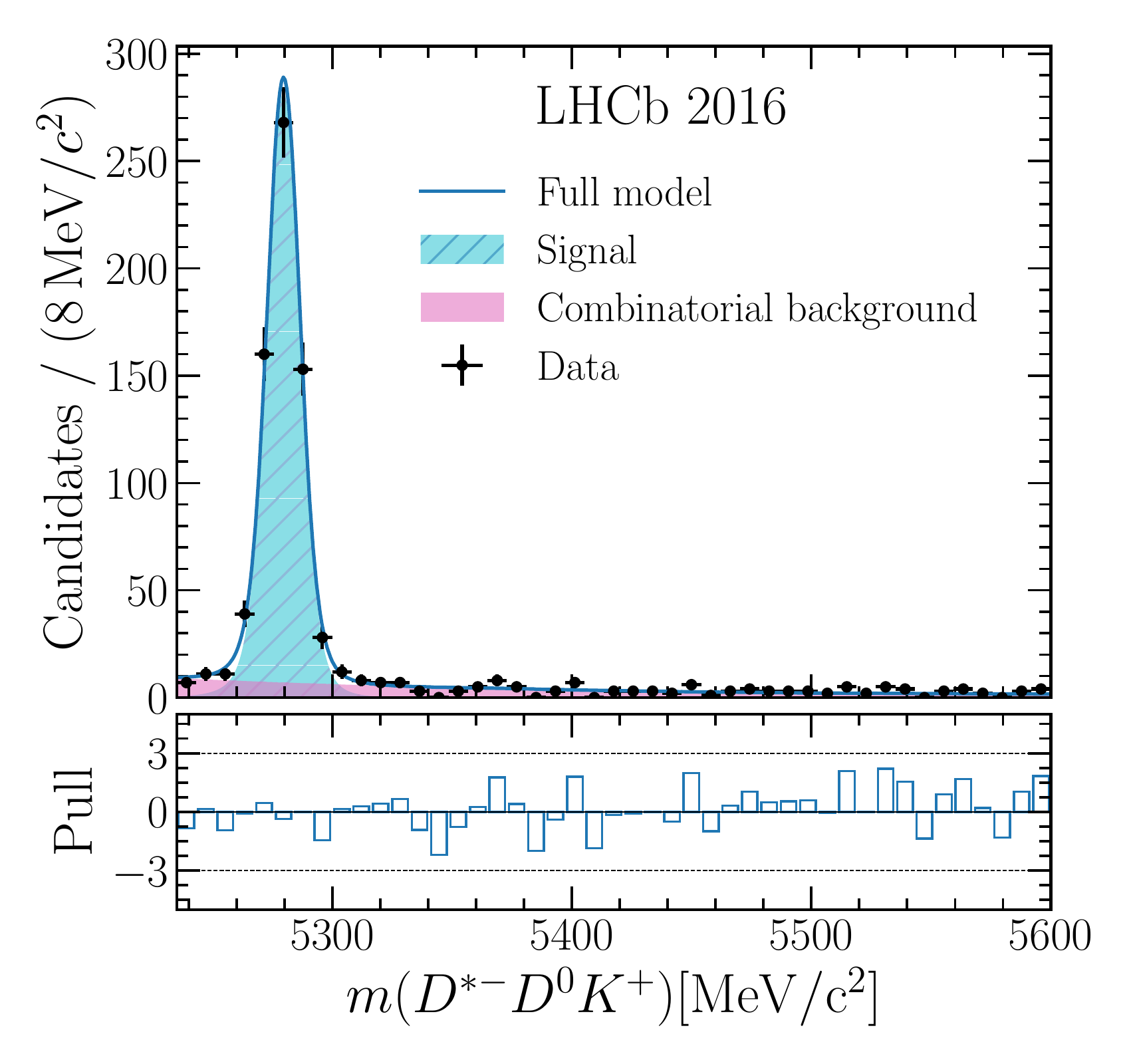}
    \end{subfigure}
    \begin{subfigure}[t]{.48\textwidth}
    \centering
    \includegraphics[width = 1.05\textwidth]{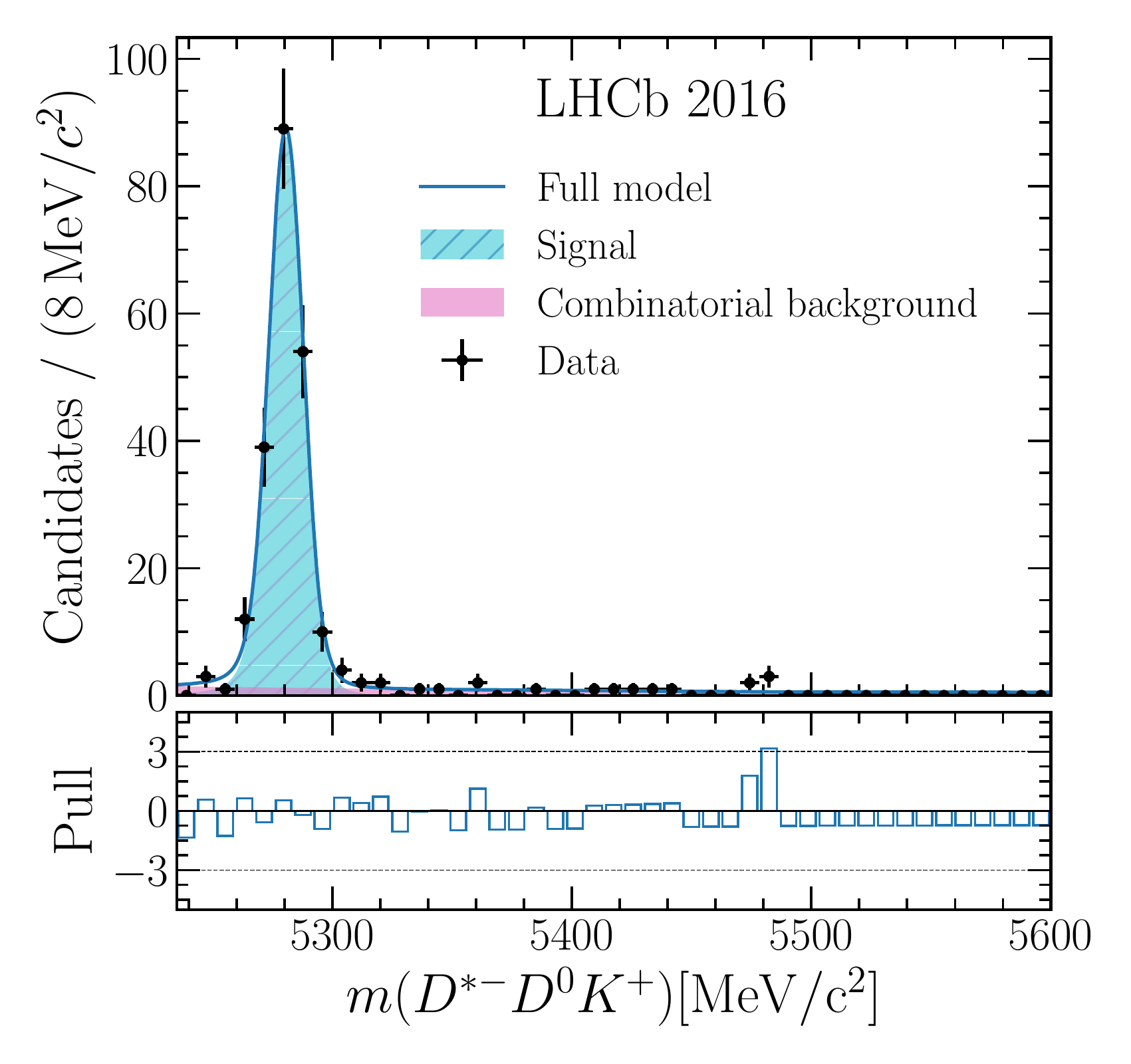}
    \end{subfigure}
  \caption{
    \small Invariant-mass distributions and fit projections for \condecay candidates, shown separately for each run period, (top) Run 1 and (bottom) 2016, and trigger category, (left) TOS and (right) TIS. The data are shown as black points with error bars and the fit components are as described in the legends. Pull projections are shown beneath each distribution.}
  \label{fig:control_fits}
\end{figure}

The invariant-mass distributions and fit projections of the fit model results for each run period and trigger category are shown in Fig.~\ref{fig:signal_fits} and Fig.~\ref{fig:control_fits} for the signal and control mode, respectively. The pulls of the fits are also shown, defined as the difference between the data and the fit with respect to the expected uncertainty. In both modes, the signal is modelled with a double-sided Crystal Ball (DSCB) function~\cite{Skwarnicki:1986xj}, defined as
\begin{equation}
    f_S(x; \Vec{\theta}) = \begin{cases}
    \big( \frac{n_L}{\lvert \alpha_L \rvert} \big)^{n_L} \exp\big(\frac{-\lvert \alpha_L \rvert ^2}{2}\big) \big( \frac{n_L}{\lvert \alpha_L \rvert} - \lvert \alpha_L \rvert -  \frac{x - \mu}{\sigma} \big) ^{-n_L}, & \text{for $\frac{x - \mu}{\sigma} \leq -\alpha_L$}\\[4pt]    
    \exp(-\frac{1}{2} \big( \frac{x-\mu}{\sigma} \big)^2), & \text{for $-\alpha_L < \frac{x - \mu}{\sigma} < \alpha_R$} \\[4pt]
    \big( \frac{n_R}{\lvert \alpha_R \rvert} \big)^{n_R} \exp\big(\frac{-\lvert \alpha_R \rvert ^2}{2}\big) \big( \frac{n_R}{\lvert \alpha_R \rvert} - \lvert \alpha_R \rvert +  \frac{x - \mu}{\sigma} \big) ^{-n_R}, & \text{for $\frac{x - \mu}{\sigma} \geq \alpha_R$,}
  \end{cases}
\end{equation}
where $\Vec{\theta}$ represents the parameters of the DSCB function, with $\mu, \sigma$ denoting the usual Gaussian shape parameters. In the control mode, the combinatorial background is modelled with an exponential function. As detailed in the paper, in the signal mode this function is modulated by a correction factor, denoted $c(x)$ such that 
\begin{equation}
    f_B(x) = c(x) \lambda e^{-\lambda x}.
\end{equation}
The full extended model is then taken as a sum of the signal and background components.

\begin{figure}[!htb]
\centering
    \begin{subfigure}[t]{.48\textwidth}
    \centering
    \includegraphics[width = 1.05\textwidth]{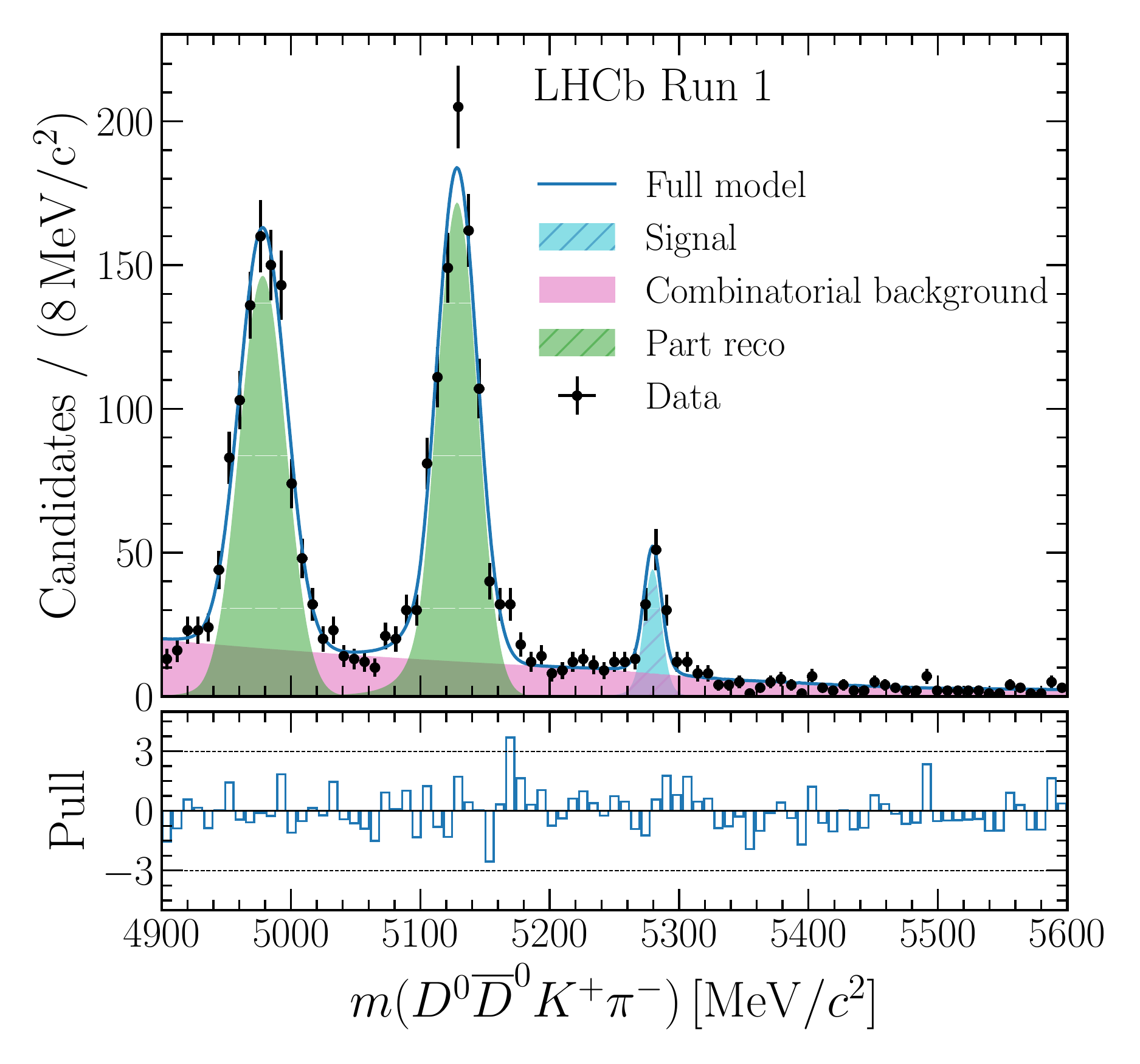}
    \end{subfigure}
    \begin{subfigure}[t]{.48\textwidth}
    \centering
    \includegraphics[width = 1.05\textwidth]{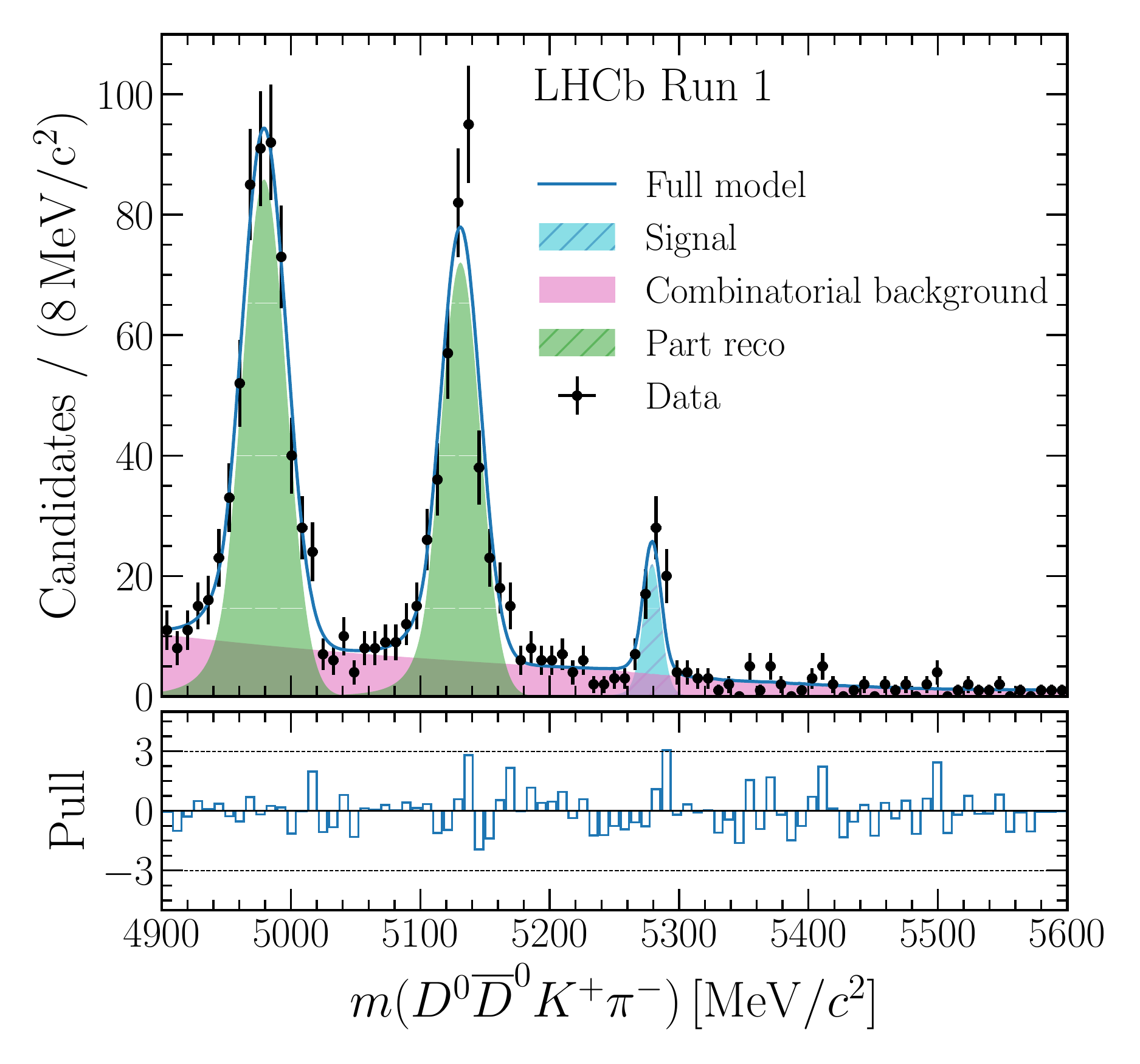}
    \end{subfigure}
    \begin{subfigure}[t]{.48\textwidth}
    \centering
    \includegraphics[width = 1.05\textwidth]{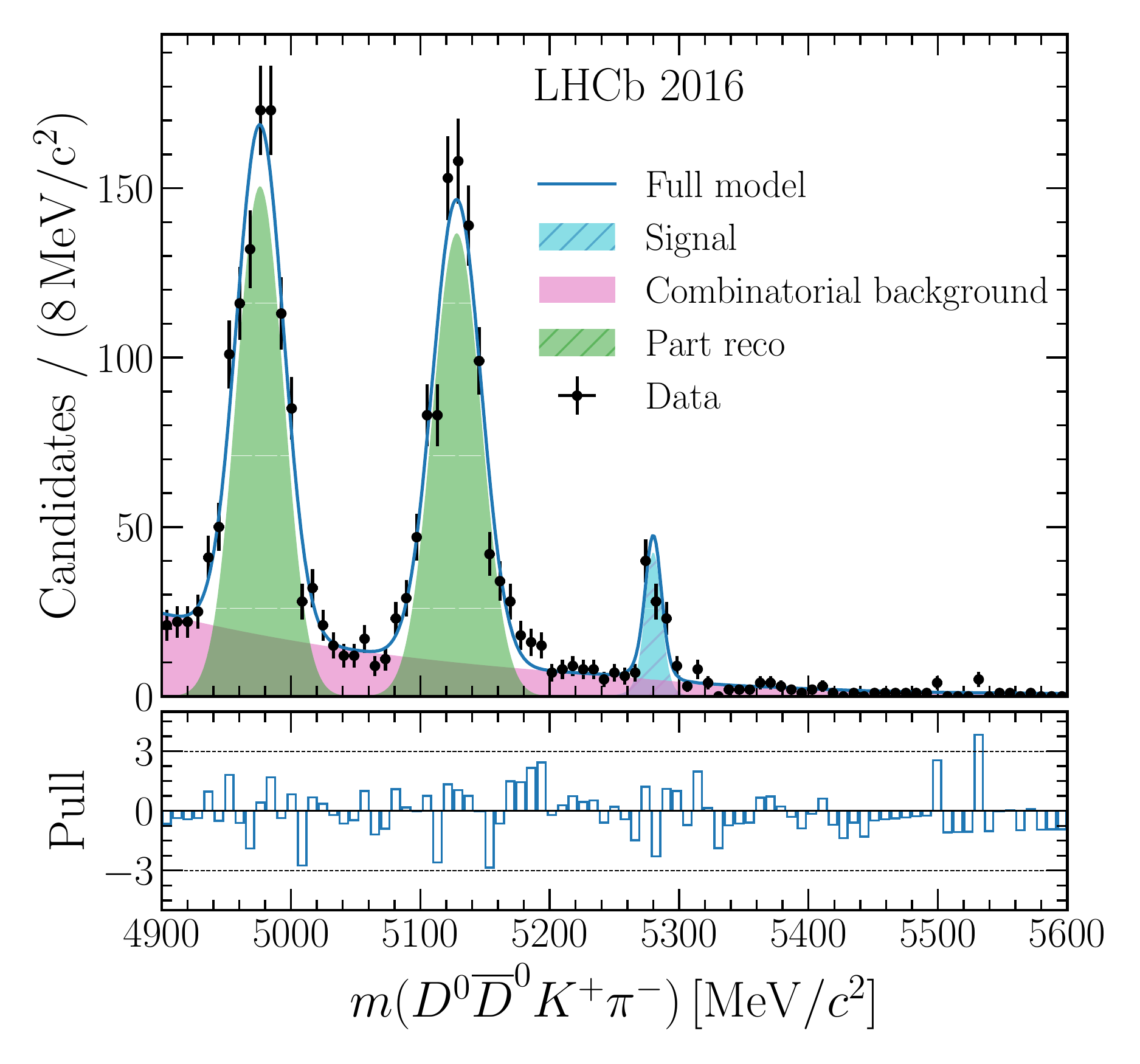}
    \end{subfigure}
    \begin{subfigure}[t]{.48\textwidth}
    \centering
    \includegraphics[width = 1.05\textwidth]{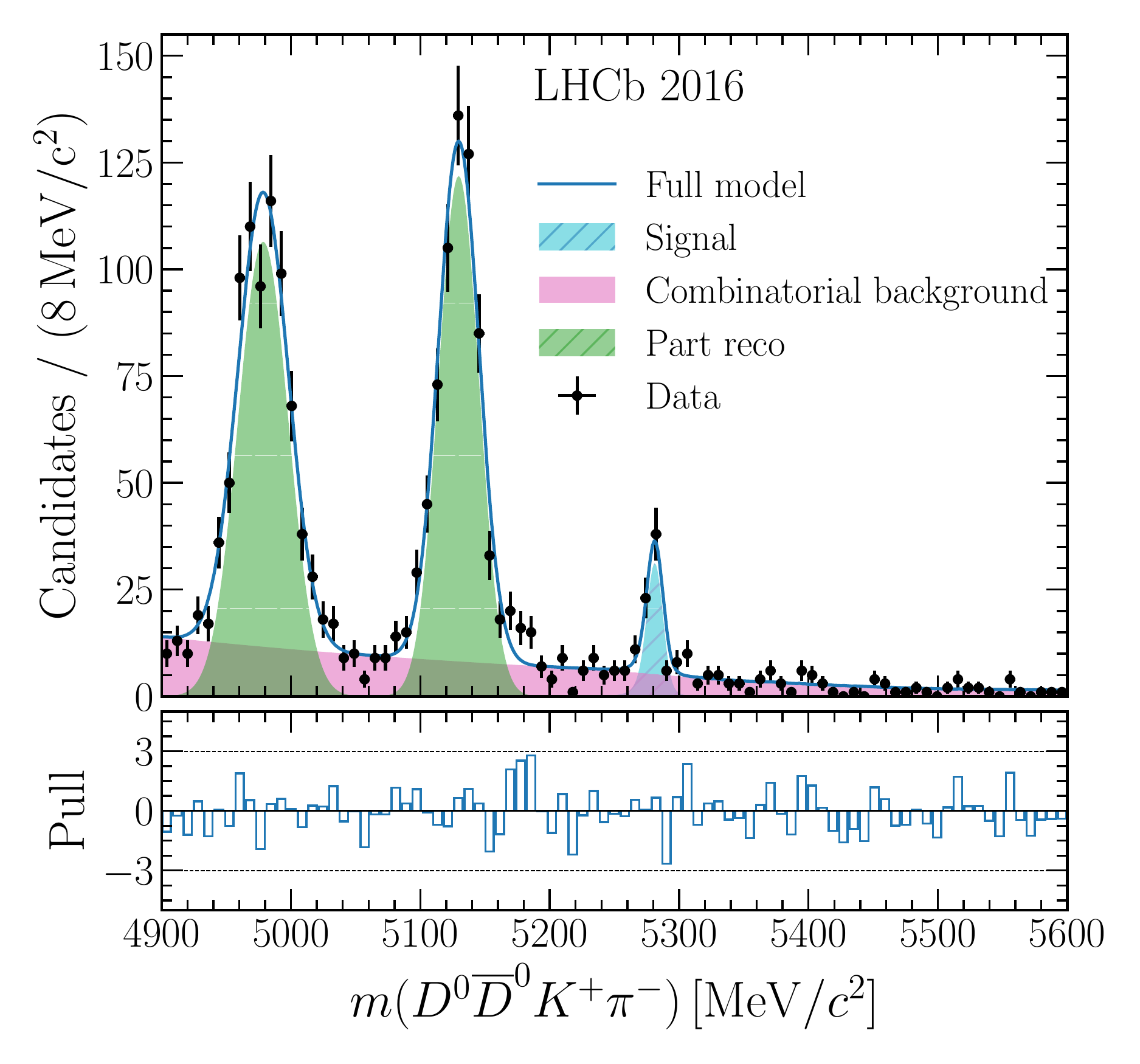}
    \end{subfigure}
  \caption{
    \small Invariant-mass distributions and fit projections for \sigdecay candidates over an extended range, shown separately for each run period, (top) Run 1 and (bottom) 2016, and trigger category, (left) TOS and (right) TIS. The data are shown as black points with error bars and the fit components are as described in the legends, where the green shaded distribution describes partially reconstructed backgrounds. Pull projections are shown beneath each distribution. The data contain a few single-charm and charmless background candidates.}
  \label{fig:part_reco}
\end{figure}

Figure~\ref{fig:part_reco} shows the invariant-mass distributions for each category in an extended range to include the partially reconstructed backgrounds $B^0 \to \Dz{}^{(*)}\Dzb{}^{(*)} \Kp \pim$ for illustrative purposes. The partially reconstructed peaks are modelled with Crystal Ball functions. Further systematic uncertainties would be required to account for the effects of modelling these additional peaks, motivating the nominal fit range \mbox{$5235 < m(\Dz \Dzb \Kp \pim) < 5600\, \mevcc$}, where the partially reconstructed contributions are negligible. As a cross check, performing the fit in this extended range results in a branching fraction ratio $\mathcal{R}$ that is in good agreement with the measurement from the nominal fit. 

\newpage
\clearpage

\pagenumbering{arabic}
\setcounter{page}{\value{zzz}+1}
\addcontentsline{toc}{section}{References}
\bibliographystyle{LHCb}
\bibliography{main,standard,LHCb-PAPER,LHCb-CONF,LHCb-DP,LHCb-TDR}
 

\newpage
\centerline
{\large\bf LHCb collaboration}
\begin
{flushleft}
\small
R.~Aaij$^{31}$,
C.~Abell{\'a}n~Beteta$^{49}$,
T.~Ackernley$^{59}$,
B.~Adeva$^{45}$,
M.~Adinolfi$^{53}$,
H.~Afsharnia$^{9}$,
C.A.~Aidala$^{82}$,
S.~Aiola$^{25}$,
Z.~Ajaltouni$^{9}$,
S.~Akar$^{64}$
% ...

{\footnotesize \it

% ...
$ ^{9}$Universit{\'e} Clermont Auvergne, CNRS/IN2P3, LPC, Clermont-Ferrand, France\\
% ...
$ ^{25}$INFN Sezione di Milano, Milano, Italy\\
% ...
$ ^{31}$Nikhef National Institute for Subatomic Physics, Amsterdam, Netherlands\\
% ...
$ ^{45}$Instituto Galego de F{\'\i}sica de Altas Enerx{\'\i}as (IGFAE), Universidade de Santiago de Compostela, Santiago de Compostela, Spain\\
% ...
$ ^{49}$Physik-Institut, Universit{\"a}t Z{\"u}rich, Z{\"u}rich, Switzerland\\
% ...
$ ^{53}$H.H. Wills Physics Laboratory, University of Bristol, Bristol, United Kingdom\\
% ...
$ ^{59}$Oliver Lodge Laboratory, University of Liverpool, Liverpool, United Kingdom\\
% ...
$ ^{64}$University of Cincinnati, Cincinnati, OH, United States\\
% ...
$ ^{67}$Syracuse University, Syracuse, NY, United States\\
% ...
$ ^{82}$University of Michigan, Ann Arbor, United States, associated to $^{67}$\\
% ...
\medskip
}
\end{flushleft}


\end{document}